\documentclass[a4paper,11pt]{article}
\pdfoutput=1 

\usepackage{pub} 

\usepackage[T1]{fontenc} 
\usepackage[italian,english]{babel}
\usepackage{hyperref}
\usepackage{ifpdf}
\usepackage{subfigure}
\usepackage{tikz}
\usetikzlibrary{arrows,shapes}
\usetikzlibrary{trees}
\usetikzlibrary{matrix,arrows} 				
\usetikzlibrary{positioning}				
\usetikzlibrary{calc,through}				
\usetikzlibrary{decorations.pathreplacing}  
\usepackage{pgffor}							

\usetikzlibrary{decorations.pathmorphing}	
\usetikzlibrary{decorations.markings}
\tikzset{
    vector/.style={decorate, decoration={snake}, draw},
    fermion/.style={draw=black, postaction={decorate}}, 
    scalar/.style={dashed,draw=black, postaction={decorate}}}
\tikzstyle{block} = [draw, rectangle, 
    minimum height=3em, minimum width=6em]

\usepackage{amssymb}
\usepackage{amsfonts}
\usepackage{epsf}
\usepackage{rotating}
\usepackage{graphicx}
\usepackage{amsmath}
\usepackage{fancyhdr}
\usepackage{lineno}
\usepackage{babel}
\usepackage{graphics}
\usepackage{pstricks}
\usepackage{color}
\usepackage{multirow}
\usepackage{float}
\newcommand{\nn}{\nonumber}
\newcommand{\lsim}{\mathrel{\mathop{\kern 0pt \rlap
  {\raise.2ex\hbox{$<$}}}
  \lower.9ex\hbox{\kern-.190em $\sim$}}}
\newcommand{\gsim}{\mathrel{\mathop{\kern 0pt \rlap
  {\raise.2ex\hbox{$>$}}}
  \lower.9ex\hbox{\kern-.190em $\sim$}}}

\newcommand{\be}{\begin{equation}}
\newcommand{\ee}{\end{equation}}
\newcommand{\bea}{\begin{eqnarray}}
\newcommand{\eea}{\end{eqnarray}}

\def\ptmiss{\not\!\!{p_T}}

\def\gph{g_{h\phi}}
\def\gev{\ensuremath{\mathrm{Ge\kern -0.1em V}}}
\def\yle{Y_{11}^\L}
\def\ylm{Y_{22}^\L}
\def\ylt{Y_{33}^\L}
\def\L{\text{\tiny L}}
\def\R{\text{\tiny R}}
\def\yre{Y_{11}^\R}
\def\yrm{Y_{22}^\R}
\def\yrt{Y_{33}^\R}

\title{\boldmath  Vacuum stability in an extended standard model with a leptoquark}

\author[a,b]{Priyotosh Bandyopadhyay}

\author[c]{and Rusa Mandal}

\affiliation[a]{Indian Institute of Technology Hyderabad, Kandi,  Sangareddy-502287, Telengana, India}
\affiliation[b]{Dipartimento di Matematica e Fisica "Ennio De Giorgi", \\ Universit\`a del Salento and INFN-Lecce, \\ Via Arnesano, 73100 Lecce, Italy}

\affiliation[c]{The 
	Institute  of Mathematical Sciences, Taramani, Chennai 600113, India \\ and \\ Homi Bhabha National Institute Training School Complex, \\ Anushakti Nagar, Mumbai 400085, India}

\emailAdd{bpriyo@iith.ac.in} 
\emailAdd{rusam@imsc.res.in}

\preprint{IITH-PH-0001/16

\hspace*{11.27cm} IMSc/2016/09/01}

\abstract{ We investigate the standard model (SM) with the extension of a charged scalar having fractional electromagnetic charge of $-1/3$ unit and with lepton and baryon number violating couplings at tree level. Without directly taking part in the electro-weak (EW) symmetry breaking, this scalar can affect stability of the EW vacuum via loop effects. The impact of such a scalar, {\it i.e.,} a leptoquark on the perturbativity of SM dimensionless couplings as well as on new physics couplings has been studied at two-loop order. The vacuum stability of the Higgs potential is checked using  one-loop renormalization group (RG) improved effective potential approach with two-loop beta function for all the couplings. From the stability analysis various bounds are drawn on parameter space by identifying the region corresponding to metastability and stability of the EW vacuum. Later we also address the Higgs mass fine-tuning issue via Veltman condition and the presence of such scalar  increases the scale up to which the theory can be considered as reasonably fine-tuned. All these constraints give a very predictive parameter space for leptoquark couplings which can be tested at present and future colliders. Especially, a leptoquark with mass $\mathcal{O}\,$(TeV) can give rise to lepton-quark flavor violating signatures via decaying into the $t\, \tau$ channel at tree level, which can be tested at the LHC or future colliders. }

\begin{document}
\maketitle
\flushbottom

\section{Introduction}
The LHC discovered a Higgs boson with mass around $125$ GeV,  which was the last keystone of standard model (SM) \cite{Aad:2012tfa,Chatrchyan:2012xdj}.  This discovery certainly proved the role of at least one scalar in electro-weak symmetry breaking (EWSB). However the experimental as well as theoretical quest for other kinds of scalars is still going on, as SM does not address many theoretical issues. One of them is the stability of vacuum at a scale above the EWSB scale. State of the art computations show that the SM vacuum is not a global minimum of the potential but its lifetime is sufficiently longer than the age of the Universe \cite{Degrassi:2012ry}. However this conclusion is very sensitive to top quark pole mass measurements and the presence of a new heavy particle, if it exists, as well. The SM alone also fails to solve the famous hierarchy problem in which the radiative corrections lead to  divergence of the SM Higgs mass.

It is true that until now no beyond the standard model particle has been discovered at colliders; however, there are certain striking discrepancies observed in the flavor sector in various experiments like LHCb, Belle, and {\it BABAR}. The discrepancies are observed mostly in rare decay modes of $B$ mesons where the SM amplitude itself is small due to loop suppression. Hence, contributions from any new heavy particle can leave a signature in various observables of these modes via loop effect and thus can hint toward indirect presence for new physics (NP). Some examples of such deviations are as follows: i) Decay $B\to D^* \tau \nu$ has $3.5 \sigma$ excess over $D^* \ell \nu$, where $\ell= e, \, \mu$, has been confirmed by all three experiments LHCb, Belle, and {\it BABAR} \cite{Lees:2012xj}--\cite{Aaij:2015yra}. ii) A lepton flavor universality ratio of $B\to K \mu^+ \mu^-$ and $B\to K e^+ e^-$ decay deviates at $2.6 \sigma$ level as observed by LHCb \cite{Aaij:2014ora}. iii) $B\to K^* \ell^+ \ell^-$ has a total $3.6 \sigma$ discrepancy in an observable $P_5^\prime$ when compared with its SM prediction \cite{Aaij:2015oid}. There are various model dependent as well as model independent explanations for the above mentioned discrepancies in the literature. The first two observations hint toward lepton non-universality physics and hence have been interest of models with leptoquark(s). Leptoquarks are a proposed particle that can decay to a lepton and a quark at tree level, and there have been previous studies on their types and properties \cite{Davidson:1993qk,Hewett:1997ce}. Recently it has been shown that the introduction of a colored, $SU(2)$ singlet, scalar leptoquark with hypercharge $-1/3$ can explain some of these anomalies \cite{Bauer:2015knc}. Such a colored scalar also successfully explains around $3\sigma$ excess in muon $g-2$ \cite{Bennett:2004pv} and can accommodate the excess of $2.4\sigma$  in a Higgs decay branching fraction to $\mu\tau$ \cite{Khachatryan:2015kon} at 8 TeV with 19.7 fb$^{-1}$ luminosity, which is, however, not seen by the ATLAS collaboration \cite{Aad:2015gha}
and needs more data to make a definite conclusion \cite{mutaua}. 

We consider the extension of SM with the scalar leptoquark which has quantum numbers $({\bf 3}, {\bf 1},-1/3)$ under SM gauge group $SU(3)_C\times SU(2)_L\times U(1)_Y$. As mentioned above the leptoquark can explain some observed anomalies \cite{Bauer:2015knc}; however, in this article, we are mainly focusing on the effect in the vacuum stability analysis of the Higgs potential in the presence of this charged scalar. The leptoquark does not participate directly in EWSB, but interestingly the presence of such a scalar affects the running of SM couplings significantly by its new `Yukawa type' couplings to leptons and quarks and also via its coupling to the Higgs boson. 

We explore the possibility of such a charged scalar coming to the rescue from metastability and/or instability of the Higgs potential for high field values. We study the behavior of the renormalization group (RG) -improved one-loop effective potential in this leptoquark model and explore the parameter space of NP couplings which cures the stability condition. In the presence of an additional scalar, the running Higgs self-coupling receives an additional positive contribution, which helps to stabilize the potential. It has been shown that, indeed, in the leptoquark model the potential is stable up to higher energy scales than the SM potential. It has also been shown that the leptoquark can modify the structure of the effective potential by introducing a new minimum which is deeper than the electroweak (EW) one and the fate of EW minimum depends on the tunneling probability between these two minima. In addition, as we demand that the theory remains valid up to the Planck scale, the NP parameters at the EW scale are constrained by the requirement that they satisfy all bounds like perturbativity, up to the Planck scale.
One of the main features of this model is that such a scalar decays via lepton-quark flavor violating couplings at tree level. Such decay final states will lead to lepton and quark flavor violating signatures at the collider, especially at the LHC. If in the near future the direct detection experiments and colliders confirm the presence of the leptoquark, our study will help to determine the stability of the EW vacuum.

The paper is organized as follows. In Sec.~\ref{model} we briefly describe the model. The beta functions for the dimensionless couplings and their perturbativity are described in Sec.~\ref{beta}. In Sec.~\ref{stab} the effect on vacuum stability analysis is discussed via RG improved effective potential approach. The tunneling probability, metastability and instability regions are also been calculated here. The Higgs mass fine-tuning has been considered in Sec.~\ref{vec} where we evaluate the Veltman condition (VC) for the model as a measure of fine-tuning.  The effect of self-coupling of the leptoquark has been discussed in Sec.~\ref{phi4} and all the results are presented in Sec~\ref{res}. Finally, in Sec.~\ref{pheno} we  discuss the phenomenological signatures that can be studied at the LHC and other colliders. Section~\ref{concl} contains some concluding remarks. 

\section{The leptoquark model}\label{model}

In this model. the SM is extended with a colored, $SU(2)_L$ singlet charged scalar $\phi$ which has $({\bf 3}, {\bf 1},-1/3)$ quantum numbers under SM gauge group. The introduction of this scalar, {\it i.e.,} the leptoquark adds additional terms in the SM Lagrangian, which is given by
\be\label{lag}
\mathcal{L}_{\phi}=(D_\mu \phi)^\dagger D^\mu \phi \,-m_\phi^2 |\phi|^2 - \,g_{h\phi}|\Phi|^2|\phi|^2 + \, \bar{Q}^c Y^{\L} i \tau_2 L \phi^*\, +\, \bar{u}^c_R Y^{\R} e_R \phi^* + h.c.
\ee
Here $\Phi$ is the SM Higgs doublet where its neutral component gets a vacuum expectation value (vev) $v$ and $Q$ and $L$ are the usual quark and lepton $SU(2)_L$ doublets defined in Eq.~\eqref{fields}.  $u^c_R$ and  $e_R$  are right-handed $SU(2)_L$ singlet up-type quark and right-handed charged lepton, respectively. 

\begin{align}\label{fields}
\Phi=\frac{1}{\sqrt{2}} \begin{pmatrix}
      0 \cr
      v+h \cr
      \end{pmatrix},\qquad
 Q=\begin{pmatrix}
     u_L \cr
     d_L \cr
     \end{pmatrix},\qquad
 L=\begin{pmatrix}
    \nu_L \cr
    e_L\cr
     \end{pmatrix}.
\end{align}

The Lagrangian in Eq.~\eqref{lag} is written in the flavor basis; however, the rotation of fermion fields should be included in the definitions of $Y^{\L,\R}$ matrices while performing the phenomenology in their mass basis. The leptoquark has electro-magnetic charge of $-1/3$ unit and is also charged under $SU(3)_C$, which makes it like the supersymmetric bottom quark. Unlike in the case of supersymmetry, here it violates the lepton and quark flavors via $Y^{\L}$ and $Y^{\R}$ couplings at tree level. On the other hand the leptoquark-Higgs coupling $g_{h\phi}$ takes part in the Higgs potential and affects the vacuum stability, which will be discussed later. In principle, one can add a renormalizable quartic coupling term $\lambda_{\phi} \phi^4$ corresponding to the leptoquark $\phi$ in Eq.~\eqref{lag}. To start with we discuss the minimal extension of SM, {\it i.e.,} without the $\lambda_{\phi} \phi^4$ term and its effect in our analysis will be discussed in Sec.~\ref{phi4} later. 

In general, the matrices $Y^{\L}$ and $Y^{\R}$ have off-diagonal terms also leading to lepton-quark flavor as well as generation-violating couplings. The of-diagonal couplings are strongly constrained by various rare meson loop induced decay modes \cite{Bauer:2015knc} and hence for most of the analysis in our paper, we will assume them to be vanishing as our results are unaffected by their small values.
For simplicity, we introduce a notation for diagonal terms in $Y^{\L}$ and $Y^{\R}$ matrices after performing the rotations for moving to the mass basis as
\begin{align}
Y^{\L,\R}\to  
\begin{pmatrix}
	 Y_{11}^{\L,\R}~~~~0 ~~~~0 \cr 
	0~~~~Y_{22}^{\L,\R}~~~~0   \cr
	0~~~~0~~ ~~Y_{33}^{\L,\R} \cr
\end{pmatrix}.
\end{align}
%
It should be noted that the main concern of this work is to study the effect of the leptoquark in EW vacuum stability, perturbativity, Higgs mass hierarchy, etc,; however, we consider the parameter space which is compatible with flavor data. There exists a strong bound on the first generation diagonal couplings $Y_{11}^{\L,\R}$ from the rare decay process $K\to \pi \nu \bar{\nu}$ \cite{Davidson:1993qk}, and thus for most of our numerical analysis, we take a conservative approach and assume them to be zero. We also emphasize that as we consider vanishing of-diagonal as well as first generation diagonal couplings, the second and third generation diagonal couplings $Y_{22}^{\L,\R}$ and $Y_{33}^{\L,\R}$ can still be taken up to $\mathcal{O}(1)$ and evade the constraints arising from other rare processes like muon $(g-2)$, $\mu \to e \gamma$, $\tau \to \mu \gamma$ \cite{Bauer:2015knc}, etc.

\begin{figure}
	\begin{center}
\mbox{\hskip -20 pt\subfigure[]{	\begin{tikzpicture}[line width=1.0pt, scale=0.8]
		\draw[scalar] (-0.2,2.3)  node[right] at (.0,2.4){$h$}-- (1,2);
		\draw[scalar] (1,2) -- (2.2,2.3) node[left] at (2,2.4){$h$};

		\draw[scalar] (2,1) node[left] at (2,1){\bf $\phi$} arc (360:2:.94);

		\draw[scalar] (2.,-0.8) node at (2.,-0.4){$h$}-- (1,0);
		\draw[scalar] (1,0) node at (-0.,-0.4){$h$}-- (-0.,-0.8);
	\end{tikzpicture}}
\hspace*{1cm}\subfigure[]{	\begin{tikzpicture}[line width=1.0pt, scale=0.8]
			\draw[scalar] (0,3)  node[right] at (0,3){$h$}-- (0,1.5);
			\draw[fermion] (0,1.5) node at (-0.6,0.0) {$t$}-- (-1,0);
			\draw[fermion] (0,1.5) -- (0.38,0.95);
			\draw[scalar] (0.38,0.95) arc (150:2:.19) node at (0.95,1){$\phi$};
			\draw[fermion] (0.38,0.95) arc (-220:2:.19) node at (0.3,0.5){$\tau$};
			\draw[fermion] (0.62,0.65) -- (1.13,-0.05) node at (0.8,0.0) {$t$};
		\end{tikzpicture}
	}}
		\caption{ We show the NP diagrams that contribute to one-loop beta functions for the SM dimensionless couplings; Higgs self-coupling $\lambda$ in (a) and top quark Yukawa coupling $y_t$ in (b).}\label{1lbf}
	\end{center}
\end{figure}
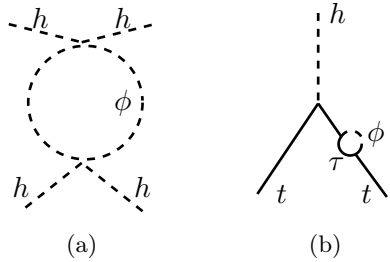

\section{One loop beta functions}\label{beta}
The behavior of various dimensionless coupling constants with energy scale $\mu$ is described in this section.  The two-loop beta functions for the dimensionless couplings are calculated for this model and the model is also implemented in {\tt SARAH-4.6.0} \cite{sarah} to calculate the running of the coupling constants via beta functions. The expressions for two-loop beta functions are too long to present here. Equations~\eqref{eq:betalam}-\eqref{eq:betag3}
show the beta functions at one-loop order for the dimensionless couplings; Higgs self-coupling $\lambda$, leptoquark-Higgs coupling $\gph$, top quark Yukawa $y_t$, new `Yukawa-type' couplings $Y_{33}^{\L,\R}$, $Y_{22}^{\L,\R}$ and $Y_{11}^{\L,\R}$, EW gauge coupling $g^\prime$ and strong gauge coupling $g_3$,  respectively. 
In Fig.~\ref{1lbf} we show the new diagrams, which arise due to the leptoquark $\phi$, contribute to one-loop running of two SM couplings $\lambda$ and $y_t$ . We can see that the Higgs self-coupling $\lambda$ is affected by leptoquark-Higgs coupling $\gph$ and we will investigate its effect 
in the stability analysis of the Higgs potential later. The leptoquark-Higgs coupling $\gph$ receives contributions from all three SM gauge couplings ($g_3$ the $SU(3)_C$ gauge coupling and $g,~ g^\prime$ the EW gauge couplings),  SM Yukawa couplings where we have only kept the largest coupling {\it i.e.,} the top quark Yukawa $y_t$. The running of $\gph$ is also affected by lepton-quark flavor violating couplings $Y^{\L,\R}_{11,\, 22,\, 33}$ as can be seen from Eq.~\eqref{eq:betagph}. Since the leptoquark $\phi$ is a $SU(2)_L$ singlet, running of $\gph$ does not get any contribution proportional to $g^4$ but due to finite hypercharge of $\phi$, $\gph$ gets contribution proportional to ${g^\prime}^4$. Top quark Yukawa coupling $y_t$ and its running are very important in SM to check the stability of the Higgs potential. The addition of the extra colored scalar $\phi$ affects the behavior of $y_t$ through its lepton-quark flavor violating couplings $Y^{\L,\R}_{11,\, 22,\, 33}$. We note that the contribution through $\gph$ is twice that of $Y^{\R}_{ii}$ for $Y^{\L}_{ii}$; this is due to the absence of right-handed neutrinos in our model and thus the couplings $Y^{\R}_{ii}$ only contributes through the leptoquark $\phi$, up type quarks and charged leptons interaction. Because of nonzero color charge and hypercharge of the leptoquark, the beta functions for the gauge couplings $g^\prime$ and $g_3$ are modified at one-loop level whereas the $SU(2)_L$ gauge coupling running remains unchanged. The expressions of one-loop beta functions are given by
\begin{align}
\label{eq:betalam}
16\pi^2 \beta^{(1)}_\lambda&= \frac{3}{4} \Big[2g^4+ \left( g^2+{g^\prime}^2\right)^2\Big]- 12 y_t^4 + \lambda  \left( 12 y_t^2-9 g^2-3 {g^\prime}^2\right) +12 \lambda^2 +6 \gph^2, \\[2ex] 
\label{eq:betagph}
16\pi^2 \beta^{(1)}_{\gph}&= \frac{{g^\prime}^4}{3} +\gph \left(6 \lambda- \frac{9}{2} g^2- \frac{13}{6} {g^\prime}^2 -8 g_3^2 \right) +4 \gph^2 +4 \gph \Big({\yle}^2 +{\ylm}^2+{\ylt}^2 \Big)  \nn \\[2ex]  
	&+2 \gph \Big( {\yre}^2 +{\yrm}^2+{\yrt}^2  \Big) + y_t^2  \Big(6 \gph - 4 {\ylt}^2 - 4 {\yrt}^2 \Big ), \\[2ex] 
\label{eq:betayt}
16\pi^2 \beta^{(1)}_{y_t}&= \frac{9}{2} y_t^3-y_t\left(\frac{9}{4}g^2+\frac{17}{12} {g^\prime}^2 +8 g_3^2 \right)+ \frac{1}{2} y_t\left({\ylt}^2+{\yrt}^2\right), \\[2ex] 
\label{eq:betaYlt}
16\pi^2 \beta^{(1)}_{\ylt}&= 4 {\ylt}^3+\ylt\left(\frac{1}{2}y_t^2-\frac{9}{2}g^2-\frac{5}{6} {g^\prime}^2 -4g_3^2 \right) +2 \ylt\left({\yle}^2 + {\ylm}^2\right) \nn \\[2ex]
	&+ \ylt\left({\yre}^2 + {\yrm}^2 +{\yrt}^2\right),  \\[2ex]
\label{eq:betaYrt}
16\pi^2 \beta^{(1)}_{\yrt}&= 3 {\yrt}^3+\yrt\left(y_t^2-\frac{13}{3} {g^\prime}^2 -4g_3^2 \right) +2 \yrt\left({\yle}^2 + {\ylm}^2 + {\ylt}^2 \right) \nn \\
 &+ \yrt\left({\yre}^2 + {\yrm}^2\right), \\[2ex]
\label{eq:betaYlm}
16\pi^2 \beta^{(1)}_{\ylm}&= 4 {\ylm}^3 - \ylm \left(\frac{9}{2}g^2+\frac{5}{6} {g^\prime}^2 +4g_3^2 \right) +2 \ylm\left({\yle}^2 + {\ylt}^2\right) \nn \\[2ex]
&+ \ylm\left({\yre}^2 + {\yrm}^2 +{\yrt}^2\right),  \\[2ex]
\label{eq:betaYrm}
16\pi^2 \beta^{(1)}_{\yrm}&= 3 {\yrm}^3- \yrm \left( \frac{13}{3} {g^\prime}^2 +4g_3^2 \right) +2 \yrm \left({\yle}^2 + {\ylm}^2 + {\ylt}^2 \right) \nn \\
&+ \yrm\left({\yre}^2 + {\yrt}^2\right), \\[2ex]
\label{eq:betaYle}
16\pi^2 \beta^{(1)}_{\yle}&= 4 {\yle}^3 - \yle \left(\frac{9}{2}g^2+\frac{5}{6} {g^\prime}^2 +4g_3^2 \right) +2 \yle\left({\ylm}^2 + {\ylt}^2\right) \nn \\[2ex]
&+ \yle\left({\yre}^2 + {\yrm}^2 +{\yrt}^2\right),  
\end{align}
\begin{align}
\label{eq:betaYre}
16\pi^2 \beta^{(1)}_{\yre}&= 3 {\yre}^3- \yre \left( \frac{13}{3} {g^\prime}^2 +4g_3^2 \right) +2 \yre \left({\yle}^2 + {\ylm}^2 + {\ylt}^2 \right) \nn \\
&+ \yre \left({\yrm}^2 + {\yrt}^2\right), \\[2ex]
\label{eq:betag1}
16\pi^2 \beta^{(1)}_{g^\prime}&= \frac{125}{18} {g^\prime}^3, \\
\label{eq:betag3}
16\pi^2 \beta^{(1)}_{g_3}&= -\frac{41}{6} g_3^3.
\end{align}

\subsection{Perturbativity }\label{pert}

\begin{figure}
\begin{center}
\mbox{\hskip -20 pt\subfigure[]{\includegraphics[width=0.53\linewidth]{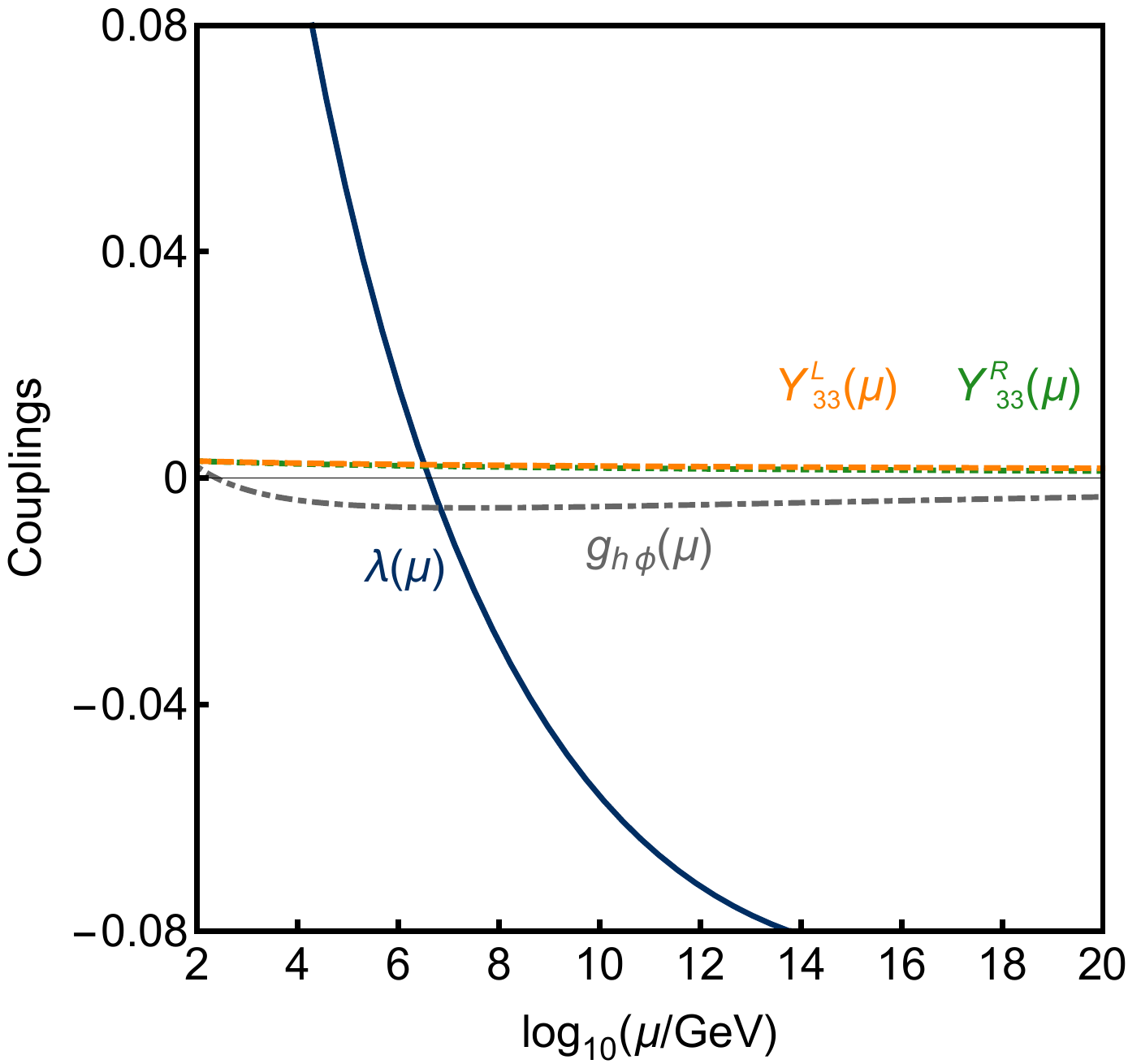}}
\subfigure[]{\includegraphics[width=0.52\linewidth]{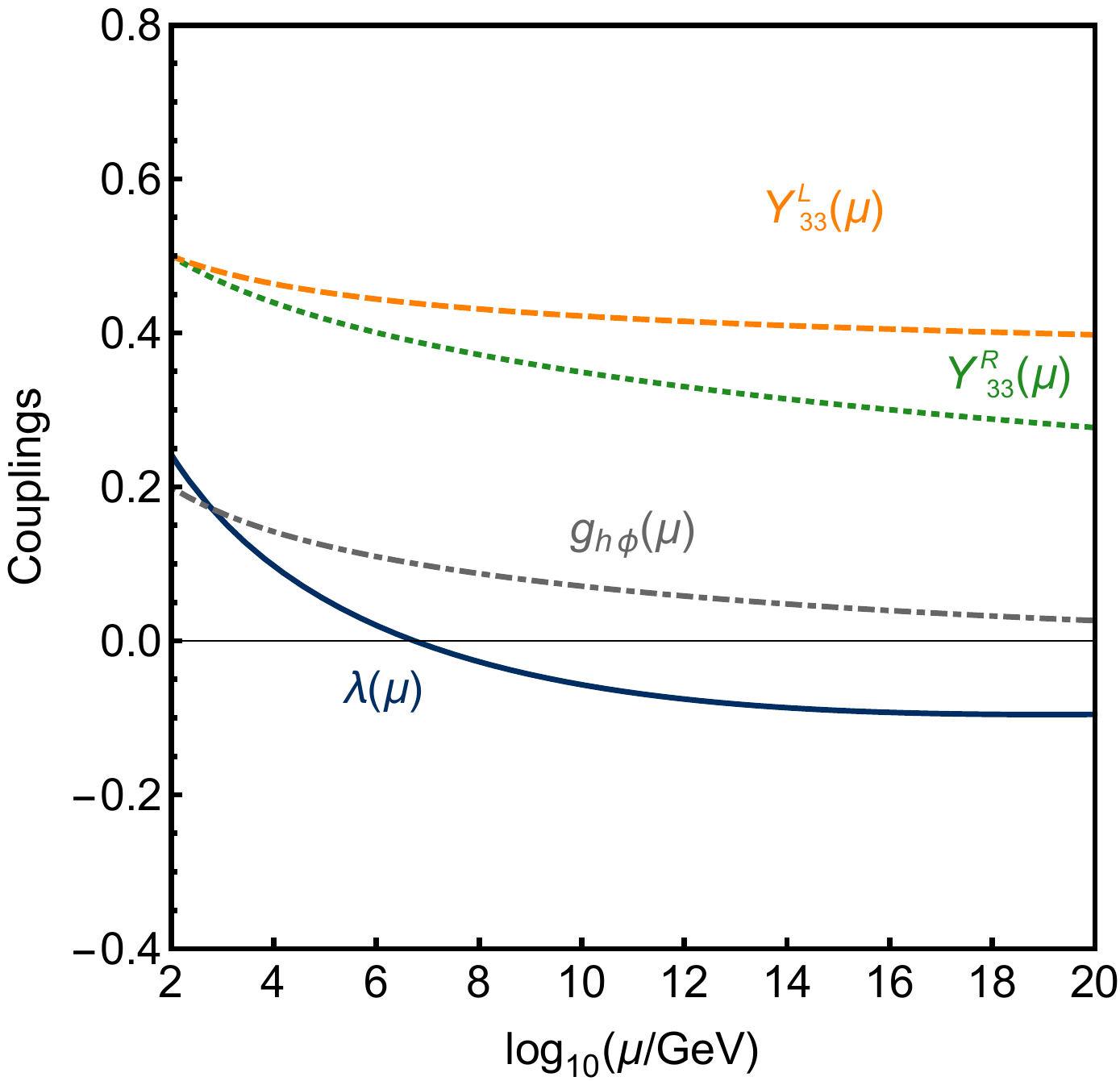}}}
\mbox{\hskip -20 pt\subfigure[]{\includegraphics[width=0.52\linewidth]{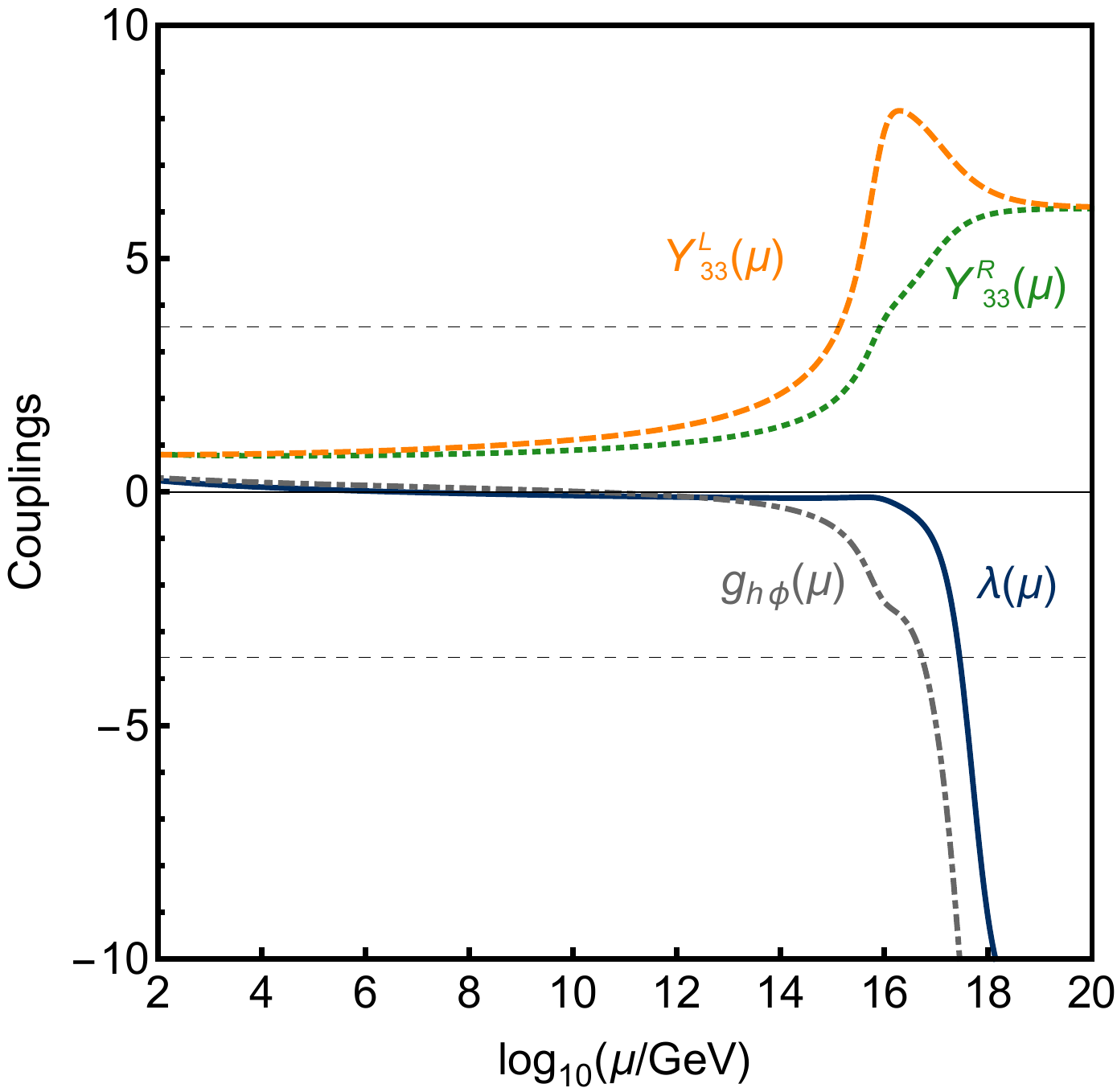}}
\subfigure[]{\includegraphics[width=0.52\linewidth]{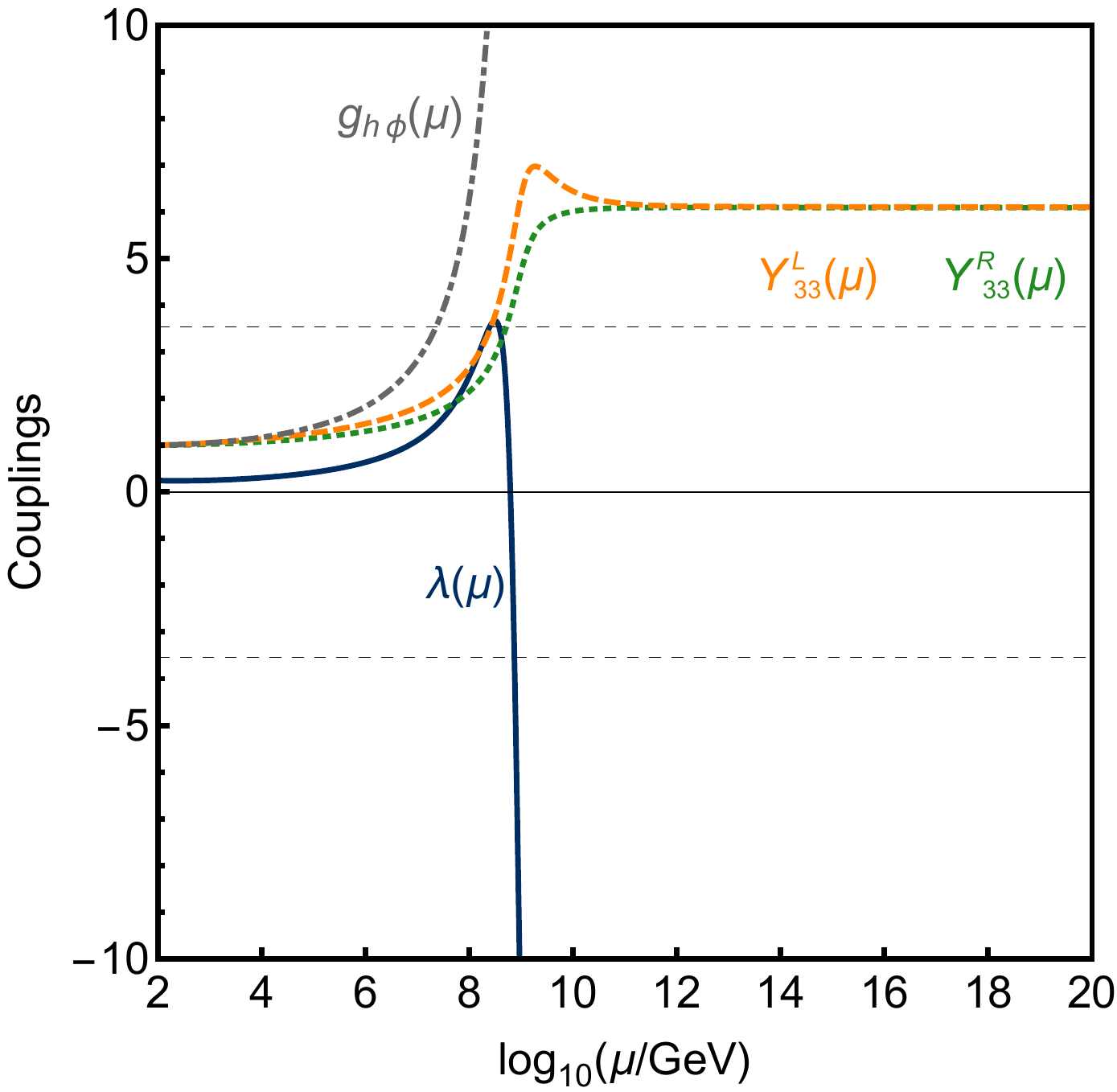}}}
\caption{The variations of two-loop coupling constants $\lambda$, $\gph$, $\ylt$ and $\yrt$ are shown with the energy scale $\mu$ in blue (solid), gray (dashed-dotted), orange (dashed) and green (dotted) curves, respectively. In the panels (a), (b), (c) and (d),  the values of NP coupling constant $\gph$, $\ylt$ and $\yrt$  at the EW scale are varied from weak ($\gph=0.002$ and $Y_{ii}^{\L,\R}=0.003$, ), moderate ($\gph=0.2$  and $Y_{ii}^{\L,\R}=0.5$), strong  ($\gph=0.3, \, Y_{ii}^{\L,\R}=0.8$), to stronger ($\gph=1, \, Y_{ii}^{\L,\R}=1$) limits, respectively, where $i=2,3$. The thin dashed lines in the last two panels denote the value $\pm \sqrt{4\pi}$, which is the perturbative limit for $\ylt$ and $\yrt$ couplings.} \label{fig:perturb}
\end{center}
\end{figure}

In this section, we study the perturbative behavior of the dimensionless couplings as we increase the validity scale of the theory. We consider all dimensionless couplings of the model are perturbative for a given value of the scale $\mu$, when the coupling constants satisfy the constraints as follows: 
\begin{eqnarray}
|\lambda |\le 4\pi, ~~|\gph |\le 4\pi ~\text{and}~ |Y_{ii}^{\L,\R}| \le \sqrt{4\pi}\,;~~i\in \{1,2,3\}.
\end{eqnarray}

Figure~\ref{fig:perturb} describes the variations of different dimensionless couplings with the energy scale $\mu$ for four different choices of parameter spaces.  The variations of two-loop coupling constants $\lambda$, $\gph$, $\ylt$ and $\yrt$ are shown with the energy scale ($\text{log}_{10}(\mu/\gev)$) in blue (solid), gray (dashed-dotted), orange (dashed) and green (dotted) curves, respectively. Figure~\ref{fig:perturb}(a) describes the weak coupling limit of the theory, where the NP couplings are very small {\it i.e.,} $\gph=0.002$ and $Y_{22}^{\L,\R}=Y_{33}^{\L,\R}=0.003$ are taken at the EW scale. The interesting feature is the scale where $\lambda$ changes sign:  around $10^{6.5}$ $\gev$, which is almost in  agreement with SM results \cite{strumia2l} in the two-loop running scenario. In the region where $\lambda$ is negative, the Higgs potential becomes unbounded from below and thus makes the theory unstable. The effect of non-negligible $\gph$ coupling at EW scale is visible in Fig.~\ref{fig:perturb}(b) where the Higgs self-coupling $\lambda$ changes sign at the $10^{6.8}$ GeV energy scale. For this plot the NP parameters at EW scale are chosen as $\gph=0.2$, $Y_{22}^{\L,\R}=Y_{33}^{\L,\R}=0.5$ and called as moderate limit of the model.
Figures~\ref{fig:perturb}(c) and \ref{fig:perturb}(d) describe the strong and stronger limits of the model defined as $\gph=0.3$, $Y_{22}^{\L,\R}=Y_{33}^{\L,\R}=0.8$ and $\gph=1$, $Y_{22}^{\L,\R}=Y_{33}^{\L,\R}=1$, respectively. In Fig.~\ref{fig:perturb}(c) $\ylt$ first hits Landau pole $\sim 10^{15}\,\gev$ and in Fig.~\ref{fig:perturb}(d) $\gph$ first hits Landau pole $\sim 10^{7}\,\gev$. Unlike the weak and moderate limits of the theory, in the strong and stronger limits ({\it i.e.,} for large  values of NP couplings at the EW scale) some of the couplings hit the Landau pole and make the theory non-perturbative and unstable. We infer from Fig.~\ref{fig:perturb} that with a smaller choice of $\gph$ and $Y_{ii}^{\L,\R}$ values at the EW scale $\gph$ evolves to negative values (see Fig.~\ref{fig:perturb}(a)), whereas for higher values of these NP couplings, $\gph$  hits the Landau pole along with the Higgs self-coupling $\lambda$ (see Fig.~\ref{fig:perturb}(d)).


\section{Vacuum stability}\label{stab}
In this section we investigate the stability of EW vacuum both at the tree level as well as with quantum corrections. For the quantum effect, we follow the RG improved effective potential approach. The effect of 
$g_{h\phi}$ is obvious at the tree level whereas the effects of other dimensionless couplings, $Y_{ii}^{\L,\R}$ can be visible via quantum loops. Below we describe our analysis.

\subsection{Effective potential}

The SM  Higgs potential at the tree level is given by 
\begin{align}
V^{\text {\tiny SM}}(h)=-\frac{1}{2}m^2 h^2+\frac{\lambda}{8} h^4,
\end{align}
where $\lambda$ and $m$ are the Higgs self-coupling and mass parameter, respectively. The scalar leptoquark $\phi$ modifies the potential at tree level via its interaction with the Higgs field as written in Eq.~\eqref{eq:V0phi}. From the renormalizable point of view one can further add a self quartic coupling for the field $\phi$, which we discuss later,
\begin{align}
\label{eq:V0phi}
V^{(0)}(h,\phi)=-\frac{1}{2} m^2 h^2+\frac{\lambda}{8} h^4 + m_\phi^2 |\phi|^2 + \frac{1}{2}\gph \,h^2 |\phi|^2.
\end{align}

In the SM, a global minimum for the Higgs potential can arise at $h \gg v$ (where $v$ is the Higgs field vev), due to the running of Higgs self-coupling $\lambda$. The large top quark Yukawa coupling, $y_t$, causes $\lambda$ to evolve to negative values at large energy scale $\mu$ as can be seen from the expressions of the beta function in Eq.~\eqref{eq:betalam}. Since the global minimum arises at a larger scale than the EW minima, we can approximate the effective potential along the $h$ axis as
\begin{align}
\label{eq:Veff}
V^{\text{eff}}\simeq \frac{\lambda^{\text{eff}}(h,\mu)}{8} h^4, ~~h \gg v.
\end{align}
The effective Higgs self-coupling $\lambda^{\text{eff}}$ is computed from the RG-improved effective potential at one-loop order. We use Landau gauge condition and $\overline{\text{MS}}$ renormalization scheme.  The one-loop correction to the RG-improved effective potential for the leptoquark model is given by,
\begin{align}
\label{eq:V1}
	V^{(1)}(h)= \sum_{i=W,Z,h,G ,t} \frac{N_i}{64\pi^2}M_i^4(h)\left[\text{ln} \frac{M_i^2(h)}{\mu^2}-C_i\right]+ \frac{3}{32\pi^2}M_{\phi}^4(h)\left[\text{ln} \frac{M_{\phi}^2(h)}{\mu^2}-\frac{3}{2}\right].
\end{align}
The first term of Eq.~\eqref{eq:V1} represents the contributions from SM particles and the second term denotes the leptoquark modification to the one-loop potential. The coefficients $N_i=6,3,1,3,-12$ and $C_i=5/6,5/6,3/2,3/2,3/2$ for the $W$ boson, $Z$ boson, Higgs particle ($h$), three Goldstone bosons ($G^\pm, G^0$) and top quark ($t$), respectively.  The mass-square expressions for the particles are given by
\begin{eqnarray}
M^2_{\scriptsize W}= \frac{1}{4}g^2(\mu)h^2,~~M^2_{\tiny{Z}}= \frac{1}{4}\left(g^2(\mu)+{g^\prime}^2(\mu)\right)h^2,~~M^2_{h}=-m^2+
\frac{3}{2}\lambda(\mu)h^2, \nn \\
M^2_{G^\pm, G^0}=-m^2+ \frac{1}{2}\lambda(\mu)h^2,~~ M^2_{t}= \frac{1}{2}y_t^2(\mu)h^2 ~~\text{and}~~M_{\phi}^2=m_\phi^2+\frac{1}{2}\gph h^2.
\end{eqnarray}
All the couplings in the above expressions are running with renormalization group equations (RGE), and the running of the Higgs field is given by $h\equiv h(\mu)= e^{\Gamma(\mu)}h_c$ where $h_c$ is the classical field and $$\Gamma(\mu)=\int_{\tiny{M_Z}}^{\mu}d\,\text{ln}(\mu^\prime)\gamma(\mu^\prime) $$
where $\gamma(\mu^\prime)$ is the Higgs field anomalous dimension.

\begin{figure}[hb]
\begin{center}
\includegraphics[width=0.55\linewidth]{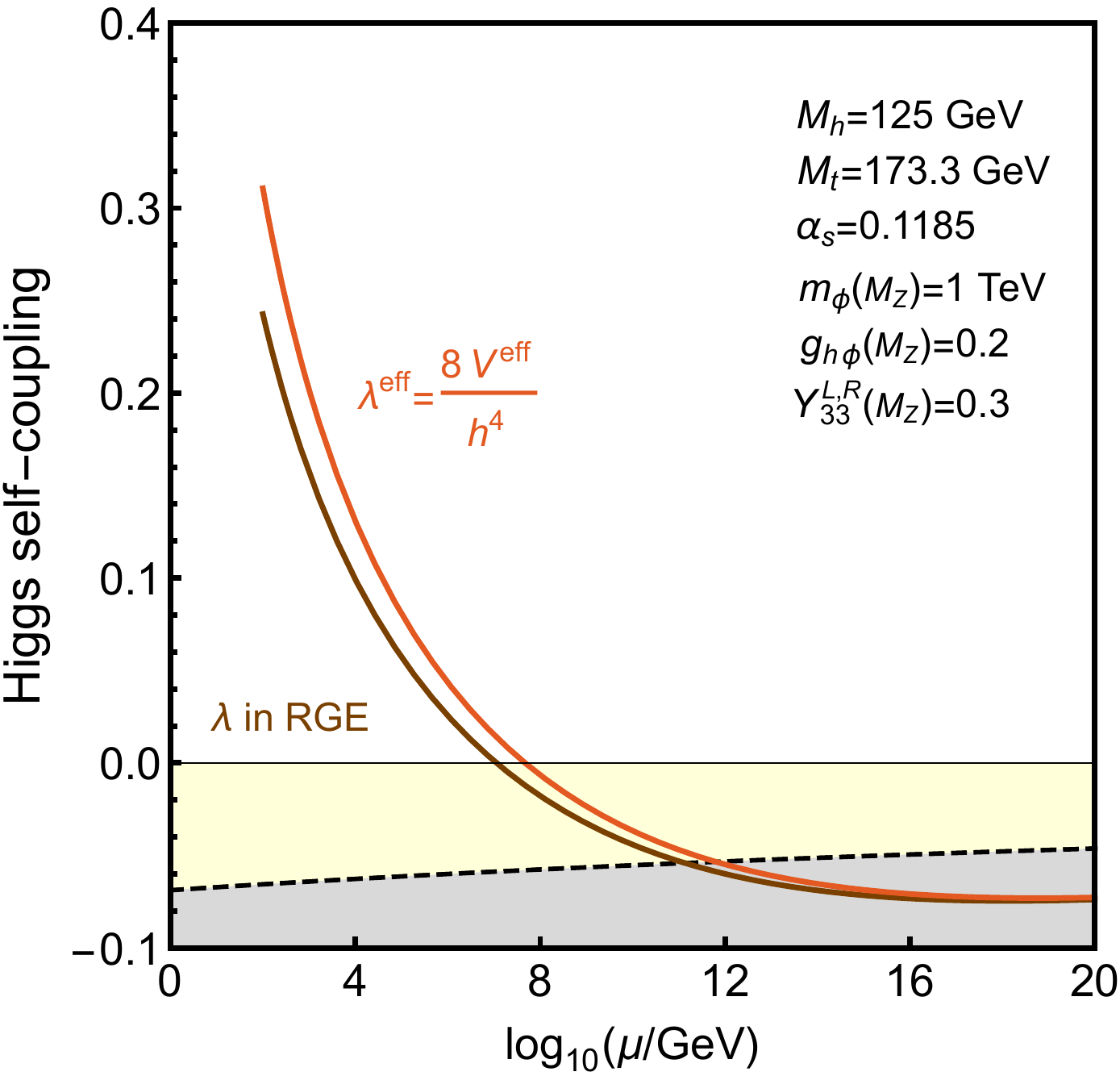}
 \caption{The comparison between the RG evolution of two-loop Higgs self-coupling $\lambda$ and the effective Higgs self-coupling $\lambda^{\text{eff}}$ with respect to the energy scale $\mu$. The parameter values used are shown in the panel. All NP couplings except $m_\phi$, $\gph$, and $Y_{33}^{\L,\R}$ are chosen to be zero at the EW scale. The yellow and gray regions correspond to the metastability and instability of the EW vacuum, respectively.} \label{fig:lambdaComp}
\end{center}
\end{figure}

Using Eq.~\eqref{eq:Veff}, the one-loop expression for $\lambda^{\text{eff}}$ is given by
\begin{align}
\label{eq:lambdaeff}
\lambda^{\text{eff}}(h,\mu) \simeq e^{4\Gamma(\mu)}\Bigg\{\lambda(\mu) &+\frac{1}{8\pi^2}\sum_{i=W,Z,h,G ,t} N_i\kappa^2_i(\mu)\left[\text{ln}\frac{\kappa_i(\mu)e^{2\Gamma(\mu)}h_c^2}{\mu^2}-C_i \right] \nn \\ &+\frac{1}{8\pi^2}\frac{3\gph^2(\mu)}{2}\left[\text{ln}\frac{\gph(\mu)e^{2\Gamma(\mu)}h_c^2}{2\mu^2}-\frac{3}{2}\right] \Bigg\},
\end{align}
where $\kappa_i$'s represent the coefficient of the $h^2$-dependent part in the mass-square expression of the corresponding particles, which are explicitly given by
\begin{eqnarray}
\kappa_{\scriptsize W}= \frac{g^2(\mu)}{4},~~\kappa_{\tiny {Z}}= \frac{g^2(\mu)+{g^\prime}^2(\mu)}{4},~~\kappa_{h}= \frac{3\lambda(\mu)}{2},~~\kappa_{G^\pm, G^0}= \frac{\lambda(\mu)}{2},~~ \kappa_{t}= \frac{y_t^2(\mu)}{2}.
\end{eqnarray}
It has been shown in Ref.~\cite{Kastening:1991gv} that the $n$-loop effective potential improved by the $(n+ 1)$-loop RGE resums all $n$th-to-leading-logarithm contributions. Hence, to ensure our calculations remain valid up to next-to-leading-logarithm approximation we will consider the beta and $\gamma$ functions of all the parameters on the rhs of Eq.~\eqref{eq:lambdaeff} to two-loop order. We point out that the whole RG-improved potential is scale independent but the one-loop approximation is not. Hence the choice of RG scale plays a crucial role in determining the behavior of the potential. It has been shown in Ref.~\cite{Casas:1994us} that a sensible selection can be $\mu=h$, where the potential remains almost scale invariant and we will stick to this choice for our numerical analysis.

It can be seen from Eq.~\eqref{eq:betalam} that the leptoquark-Higgs coupling $\gph$ has a positive contribution to the beta function of $\lambda$, irrespective of the sign of $\gph$. Thus the effect of $\gph$ increases the scale up to which $\lambda^{\text{eff}}$ remains positive or in other words the Higgs potential remains stable. Fig.~\ref{fig:lambdaComp} shows a comparison between the RG evolution of two-loop Higgs self-coupling $\lambda$ and the effective Higgs self-coupling $\lambda^{\text{eff}}$ with respect to  the energy scale $\mu$. The parameter values used to obtain the plot are shown in the panel. All NP couplings except $m_\phi$, $\gph$ and $Y_{33}^{\L,\R}$ are chosen to be zero at the EW scale. We want to point out that the running of $\lambda$ is insensitive to $Y_{ii}^{\L,\R}(M_Z)$ values as they contribute only at two-loop level. Hence for all the numerical analysis in this section, we assume $Y_{11,22}^{\L,\R}(M_Z)$ are zero and  the third generation couplings $Y_{33}^{\L,\R}$ are finite at the EW scale for simplicity. However for the perturbativity analysis in Sec.~\ref{pert}, the RG runnings of $\gph$, $\ylt$ and $\yrt$ couplings are sensitive to $Y_{11,22}^{\L,\R}(M_Z)$ values even at one-loop level (Eqs.~\eqref{eq:betagph}--\eqref{eq:betaYre}) and thus have been taken finite. The finite values of $Y_{11,22}^{\L,\R}(M_Z)$ do not change the results obtained in stability analysis.
It can be seen from Fig.~\ref{fig:lambdaComp} that there is one order difference between the zero crossings of $\lambda$ and $\lambda^{\text{eff}}$. We note that actually the potential maximizes at the point where $\lambda^{\text{eff}}\sim 0$ \cite{Casas:1994qy}. The yellow and the gray regions correspond to the metastability and instability of the vacuum and will be discussed in more detail in the next section.
%

\subsection{Tunneling probability and  metastability}
\begin{figure}[ht]
\begin{center}
\includegraphics[width=0.53\linewidth]{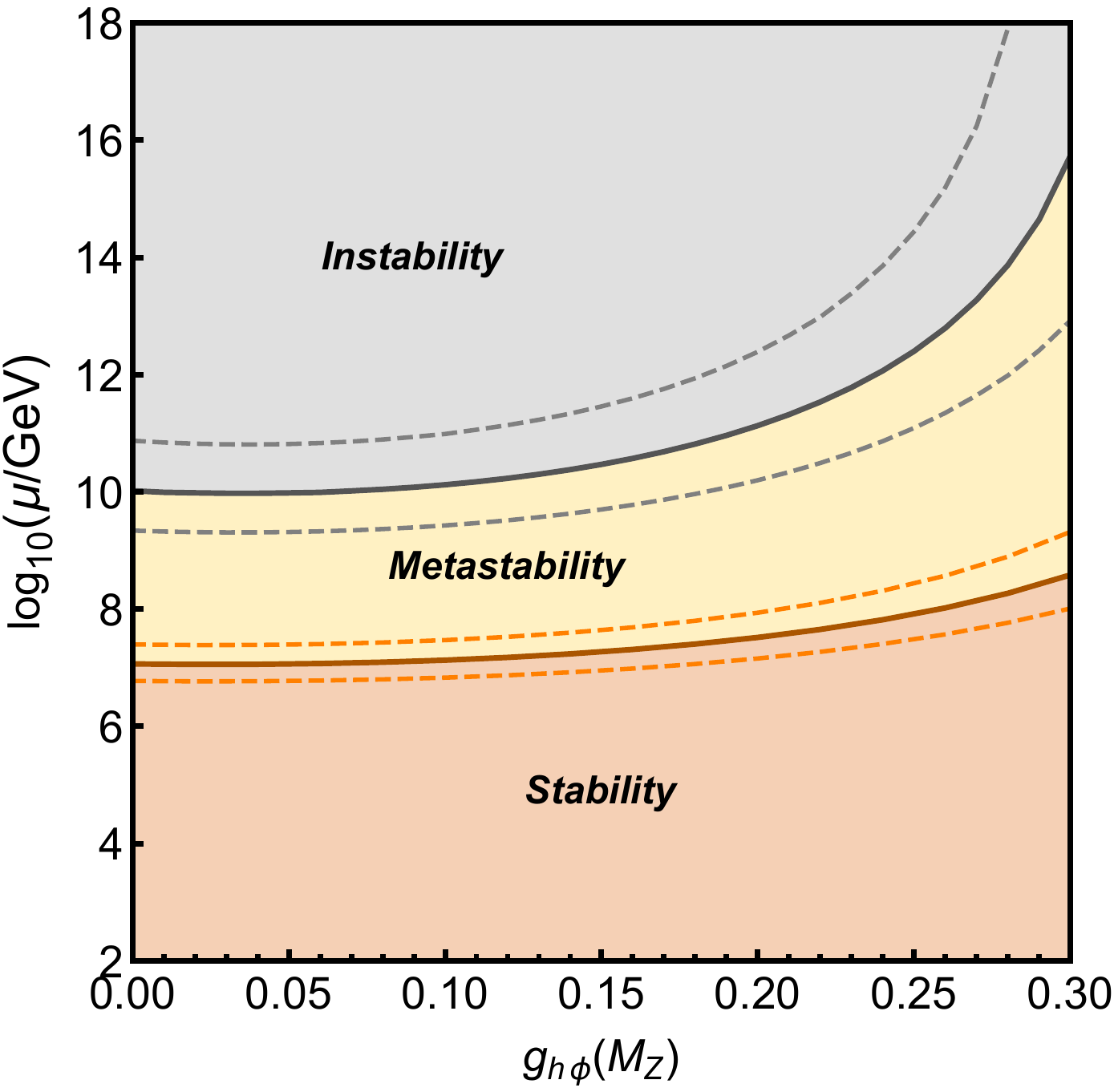}
 \caption{The stability, metastability and instability regions for EW vacuum are shown as a function of the leptoquark-Higgs coupling constant $\gph$ value at the EW scale. The light brown region (stability) below the brown solid curve assures the effective Higgs self-coupling $\lambda^{\text{eff}}(\mu) > 0$. The yellow region (metastability) denotes $\lambda^{\text{eff}}(\mu) > \frac{-0.065}{1-0.01\,\text{ln}(v/\mu)}$ and the gray region (instability) corresponds to $\lambda^{\text{eff}}(\mu) < \frac{-0.065}{1-0.01\,\text{ln}(v/\mu)}$. The dashed lines show the uncertainty due to top quark mass $M_t=173.3 \pm 0.87\, \gev$ in the corresponding cases \cite{mtop}. The plot is obtained for $\alpha_s = 0.1185$ and NP coupling $Y_{33}^{\L,\R}=0.4$. For $\gph > 0.3$ the vacuum is (meta)stable up to the Planck scale.}\label{fig:stability}
\end{center}
\end{figure}

It is known that for $\lambda^{\text{eff}} \lesssim 0$ the potential becomes unbounded from below which imposes a 
direct bound to the validity scale of the model.  As mentioned in the previous section, along the $h$ axis, the Higgs self-interaction term dominates, and hence in the presence of a leptoquark quartic coupling term ($\lambda_\phi \phi^4$ with $\lambda_\phi>0$), negative values of leptoquark-Higgs coupling $\gph$ are also allowed for large Higgs field values, and the potential still remains bounded from below from both the field directions. But in the absence of leptoquark quartic coupling, the value of $\gph$ is not restricted and any arbitrary large negative value (within the perturbativity regime) can make the potential unbounded from the leptoquark field direction. Hence, to avoid such cases, we will assume $\gph$ is positive at the EW scale. The effect of $\gph$ on the running of $\lambda $ leads to new bounds on the validity scale of the model. There are three different classifications of the EW vacuum depending upon the running of Higgs self-coupling as demonstrated in Fig.~\ref{fig:stability} and discussed below.

{\bf Stability bound:} The region where the EW minima of the scalar potential is the global minimum and the Higgs self-coupling $\lambda^{\text{eff}} >0$, is known as stability region. The variation of the scale up to which $\lambda^{\text{eff}} >0$, with the leptoquark-Higgs coupling $\gph$ value at EW scale is shown in the light brown region in Fig.~\ref{fig:stability}. The region corresponds the stability region of the potential with the solid brown curve denoting the upper bound, which is obtained by equating $\lambda^{\text{eff}}=0$ with top quark mass $M_t=173.3\,\gev$ and strong coupling constant $ \alpha_s = 0.1185$. The orange dashed lines denote the variation of the stability bound due to top quark mass uncertainty $M_t=173.3 \pm 0.87\, \gev$ \cite{mtop}. \\ \\

{\bf Metastability bound:} In the case in which there exists a global minimum other than the EW vacuum, one  needs to examine the stability of the EW vacuum by calculating the decay rate of the EW vacuum to that global minimum and ensuring that the rate is greater than the age of the Universe. The probability of tunneling to the deeper vacuum is given by \cite{strumia2l},
\begin{align}
p= T_U^4 \mu^4 e^{-\frac{8\pi^2}{3 |\lambda^{\text{eff}}(\mu)|}},
\end{align}
where $T_U$ is the age of the Universe and $\mu$ denotes the scale where the probability is maximized.
Assuming $T_U\sim 10^{10}$ yr and requiring that the EW vacuum tunneling lifetime is always higher than the lifetime of the Universe, the condition \cite{strumia2l}
\begin{align}
\lambda^{\text{eff}}(\mu) > \frac{-0.065}{1-0.01\,\text{ln}(v/\mu)}
\end{align}
arises, where $v$ is the Higgs vev.
The region where $\frac{-0.065}{1-0.01\,\text{ln}(v/\mu)}< \lambda^{\text{eff}} <0$  is termed the metastable region and is shown as the yellow region in Fig.~\ref{fig:stability}. We see that for $M_t=173.3~ \gev$ the metastable region is accessible until the grand unified theory (GUT) scale, {\it i.e,} $\sim 10^{15}$ GeV for $\gph \sim 0.3$. While considering the top quark mass uncertainty the accessibility of the metastable region is enhanced to the
Planck scale for $\gph(M_Z) > 0.3$.\\\\

{\bf Instability bound:} The scalar potential becomes unstable for large negative values of $\lambda^{\text{eff}}$. In this case, there exists a deeper global minimum other than the EW one, and the EW vacuum eventually decays to the global minimum within the lifetime of our universe. It was shown in Ref.~\cite{strumia2l} that this case arises for $\lambda^{\text{eff}}(\mu) < \frac{-0.065}{1-0.01\,\text{ln}(v/\mu)}$, and the corresponding region is highlighted as the gray region in Fig.~\ref{fig:stability}. The lower bound of the instability scale is shown as a gray solid curve, which is obtained by equating $\lambda^{\text{eff}}(\mu) = \frac{-0.065}{1-0.01\,\text{ln}(v/\mu)}$ with top quark mass $M_t=173.3~\gev$ and strong coupling constant $ \alpha_s = 0.1185$. The variation due to top quark mass uncertainty is depicted by gray dashed lines.

From the parameter space of the metastable region we have considered two particular benchmark points for which the global minima of the Higgs potential occur around the Planck scale as given below:
\begin{center}
{\bf BP1}: $m_\phi=1\, \text{TeV}$, $\gph=0.4$, $Y_{33}^{\L,\R}=0.4$,\\

{\bf BP2}: $m_\phi=1 \,\text{TeV}$, $\gph=0.5$, $Y_{33}^{\L,\R}=0.5$. 
\end{center}

\begin{figure}[H]
	\begin{center}
		\mbox{\hskip -20 pt\subfigure[]{\includegraphics[width=0.53\linewidth]{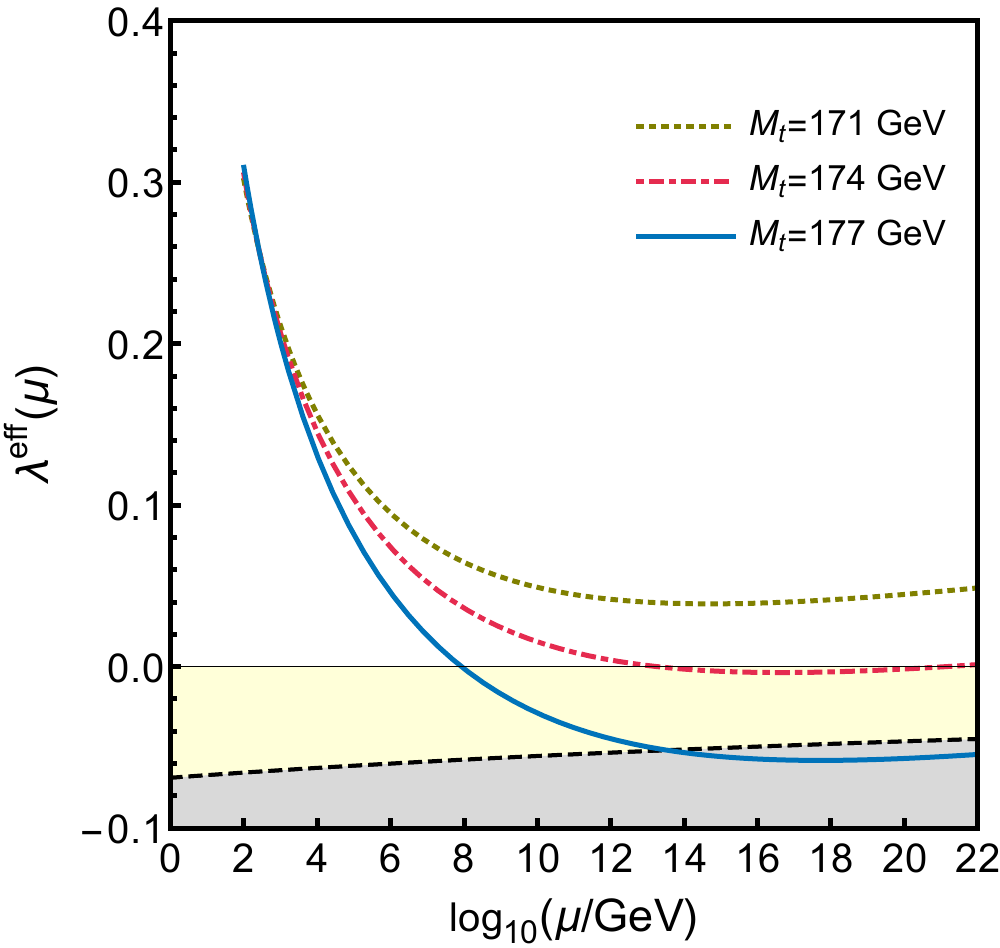}}
			\subfigure[]{\includegraphics[width=0.53\linewidth]{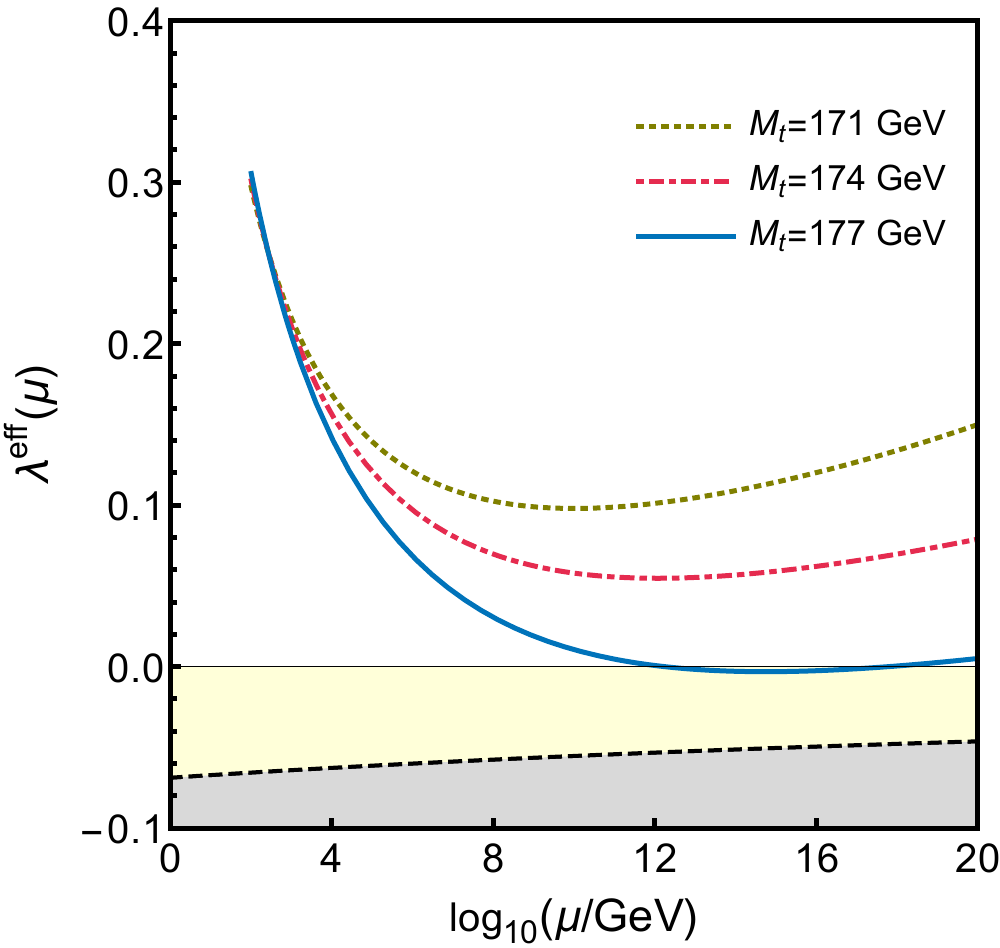}}}
		\mbox{\hskip -20 pt\subfigure[]{\includegraphics[width=0.53\linewidth]{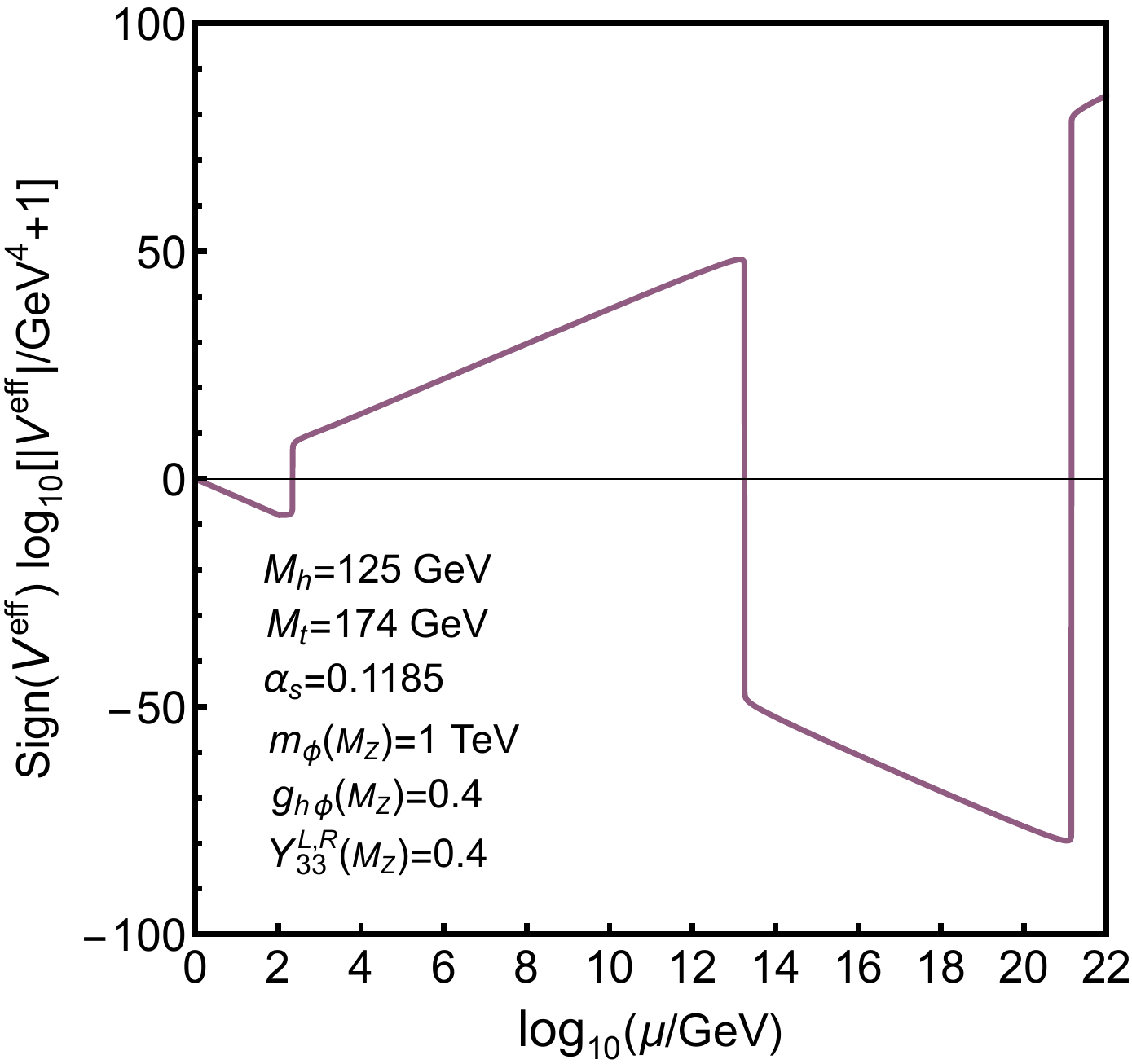}}
			\subfigure[]{\includegraphics[width=0.53\linewidth]{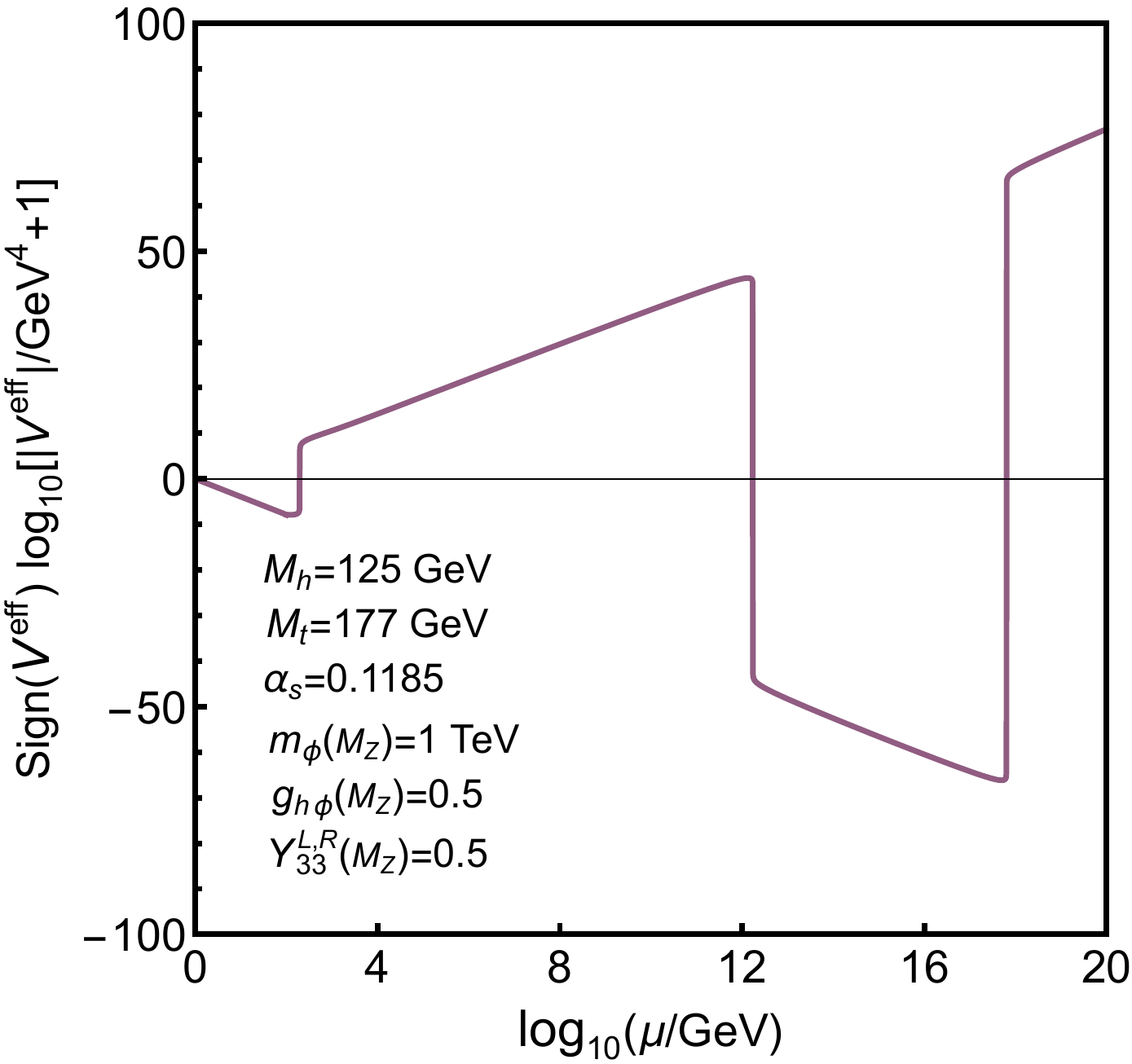}}}
		\caption{The variations of effective Higgs self-coupling $\lambda^{\text{eff}}$ as a function of energy scale $\mu$ for top quark mass $M_t=171,\, 174,\, 177$ GeV are presented in panels (a) and (b), and corresponding effective potentials (for a particular choice of $M_t$) are shown in panels (c) and (d). For panels (a) and (c) NP parameters are $m_\phi=1\, \text{TeV}$, $\gph=0.4$ and $Y_{33}^{\L,\R}=0.4$ (BP1), and for panels (b) and (d) the corresponding values are $m_\phi=1 \,\text{TeV}$, $\gph=0.5$ and $Y_{33}^{\L,\R}=0.5$ (BP2). All other NP couplings are chosen zero at the EW scale. The yellow and gray regions shown in panels (a) and (b)  correspond to metastability and instability of the EW vacuum, respectively. Both the EW and global minima of the potential for the two benchmark points can be seen from panels (c) and (d) (see the text for details).} \label{fig:meta}
	\end{center}
\end{figure}

The results are highlighted in Fig.~\ref{fig:meta}. The effective Higgs self-coupling $\lambda^{\text{eff}}$ is plotted in Figs.~\ref{fig:meta}\,(a) and \ref{fig:meta}\,(b) as a function of energy scale $\mu$ for the above-mentioned benchmark points. The corresponding effective potentials (for a particular choice of $M_t$) are shown in Figs.~\ref{fig:meta}\,(c) and \ref{fig:meta}\,(d), respectively. The variations due to top quark mass ($M_t$) are also visible. Fig.~\ref{fig:meta}\,(a) and Fig.~\ref{fig:meta}\,(c) are obtained for NP parameters $m_\phi=1\, \text{TeV}$, $\gph=0.4$ and $Y_{33}^{\L,\R}=0.4$ (BP1) and  Fig.~\ref{fig:meta}\,(b) and Fig.~\ref{fig:meta}\,(d) correspond to $m_\phi=1 \,\text{TeV}$, $\gph=0.5$ and $Y_{33}^{\L,\R}=0.5$ (BP2) at the EW scale. 
The $y$ axis plotted in Figs.~\ref{fig:meta}(c) and \ref{fig:meta}(d) is given by \cite{Casas:1994qy}
\be
\rm{Sign}(V_{\rm{eff}})log_{10}[|V_{\rm{eff}}|/\rm{GeV}^4+1].
\ee
This choice has been made to give a continuous picture of the shape of the potential starting from the EW to Planck scale and the two minima (EW and global) can be realized in the same plot \cite{Casas:1994qy}.
It can be seen that the maximum of the potential (within the Planck scale) and the new minimum of the potential arise at the two zero crossings of $\lambda^{\text{eff}}$, respectively. 
For the first benchmark point (BP1), the red (dashed) curve crosses the $x$ axis at a scale $\mu\sim 10^{13}\, \gev$ where the potential is at maximum and after it, potential goes to negative values, developing a new minima at the second zero crossing of 
$\lambda^{\text{eff}}$ around $\mu\sim 10^{21}\,\gev$, which is beyond the Planck scale. One can easily see that the new minimum is much deeper than the EW one and so is a global minimum. However, tunneling probability calculation ensures that the decay time to this global minimum is larger than the age of Universe as the running of $\lambda^{\text{eff}}$ remains within the metastable bound. The second benchmark point (BP2) also highlights a similar kind of situation, except for this case, the global minimum occurs before the Planck scale around $\mu\sim 10^{18}\,\gev$.  
We notice that such behavior is very sensitive to the top quark mass and a higher top quark mass can lead to instability or metastability, as can be seen from both Figs.~\ref{fig:meta}\,(a) and \ref{fig:meta}(b).

\section{Fine-tuning and Veltman condition}\label{vec}

In an effective field theory approach with an ultraviolet cutoff $\Lambda$, the Higgs self-energy receives quadratically divergent corrections from loop diagrams such that
\begin{align}
M_h^2={(M_h^2)}_{bare} + \mathcal{O}(\lambda,\,g_3^2,\,g^2\!,\,{g^\prime}^2\!,\,y_t) \Lambda^2,
\end{align}
where ${(M_h^2)}_{bare}= v^2 \lambda$ the bare Higgs mass, $\lambda$ is the Higgs self-coupling; $g_3,~g$ and $g^\prime$ are the three SM renormalized gauge couplings; and $y_t$ is the top quark Yukawa coupling constant. It was shown by Veltman \cite{vltmn} that the one-loop correction to the Higgs mass within the SM is 
\begin{figure}[hb]
\begin{center}
\includegraphics[width=0.5\linewidth]{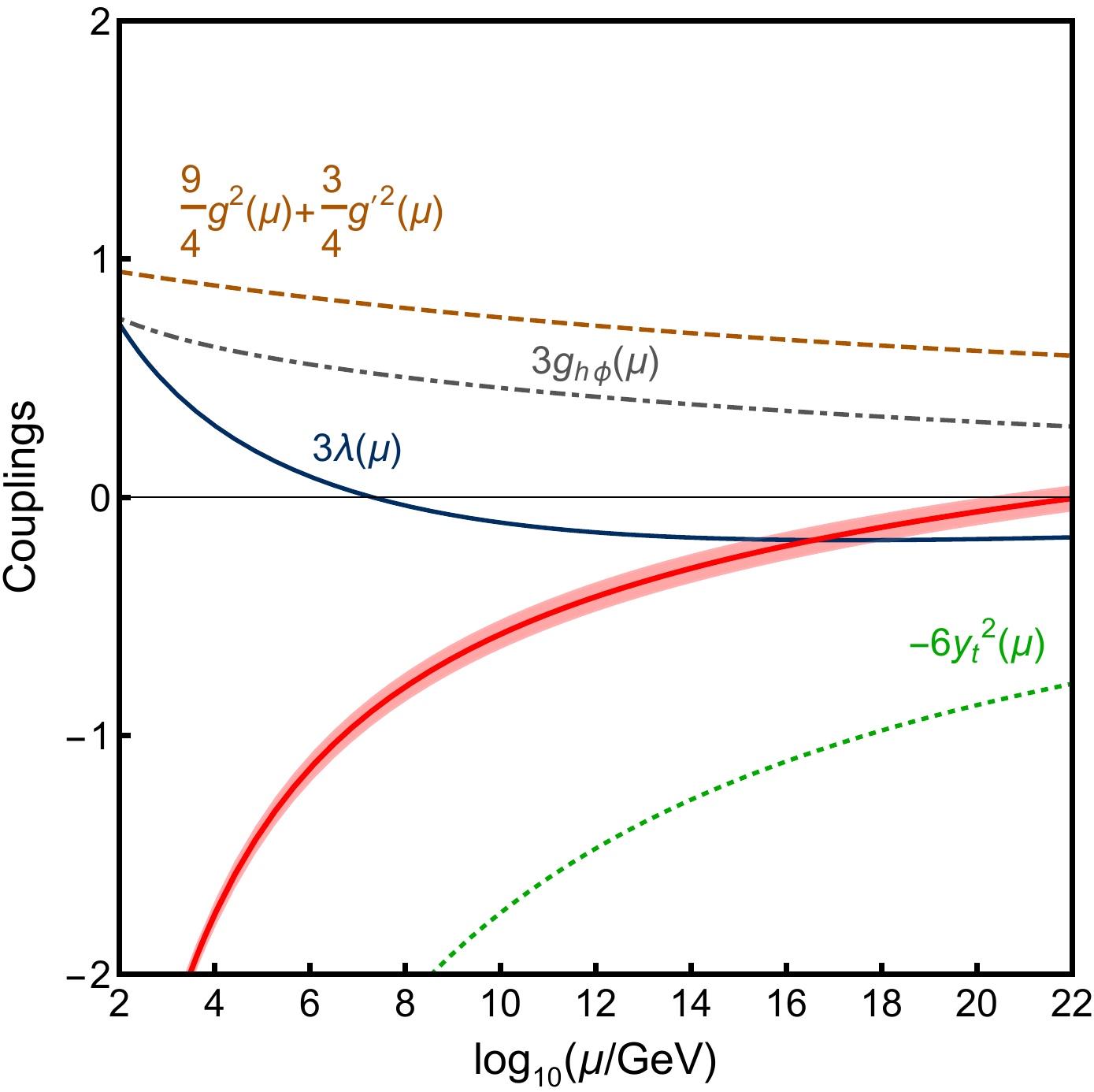}
\caption{The running of the VC (red curve) with the renormalization scale $\mu$. All contributions to Eq.~\eqref{VClq2} from different couplings are shown also. The plot is obtained by choosing: $M_h=125\,\gev$ for the Higgs mass and $\gph=0.25$ for leptoquark-Higgs coupling constant and all other NP couplings are vanishing at the EW scale. The red band corresponds to the variation of the VC due to the uncertainty in top quark mass $M_t=173.3 \pm 0.87\, \gev$ \cite{mtop}. The VC is satisfied beyond the Planck scale for this set of parameter space.} \label{fig:velt}
 \end{center}
\end{figure}
\be\label{VC}
\delta M^2_h= \frac{\Lambda^2}{16 \pi^2}\left( 3\lambda + \frac{9}{4} g^2 + \frac{3}{4} g'^2 - 6 y^2_t \right).
\ee

The absence of quadratic divergences in the Higgs mass can only be possible if cancellation occurs between fermionic and bosonic contributions, and such a demand is known as the VC which is given by at one-loop order as
\be\label{VC2}
 3\lambda + \frac{9}{4} g^2 + \frac{3}{4} g'^2 - 6 y^2_t=0.
\ee

\begin{figure}[b]
	\begin{center}
		\includegraphics[width=0.5\linewidth]{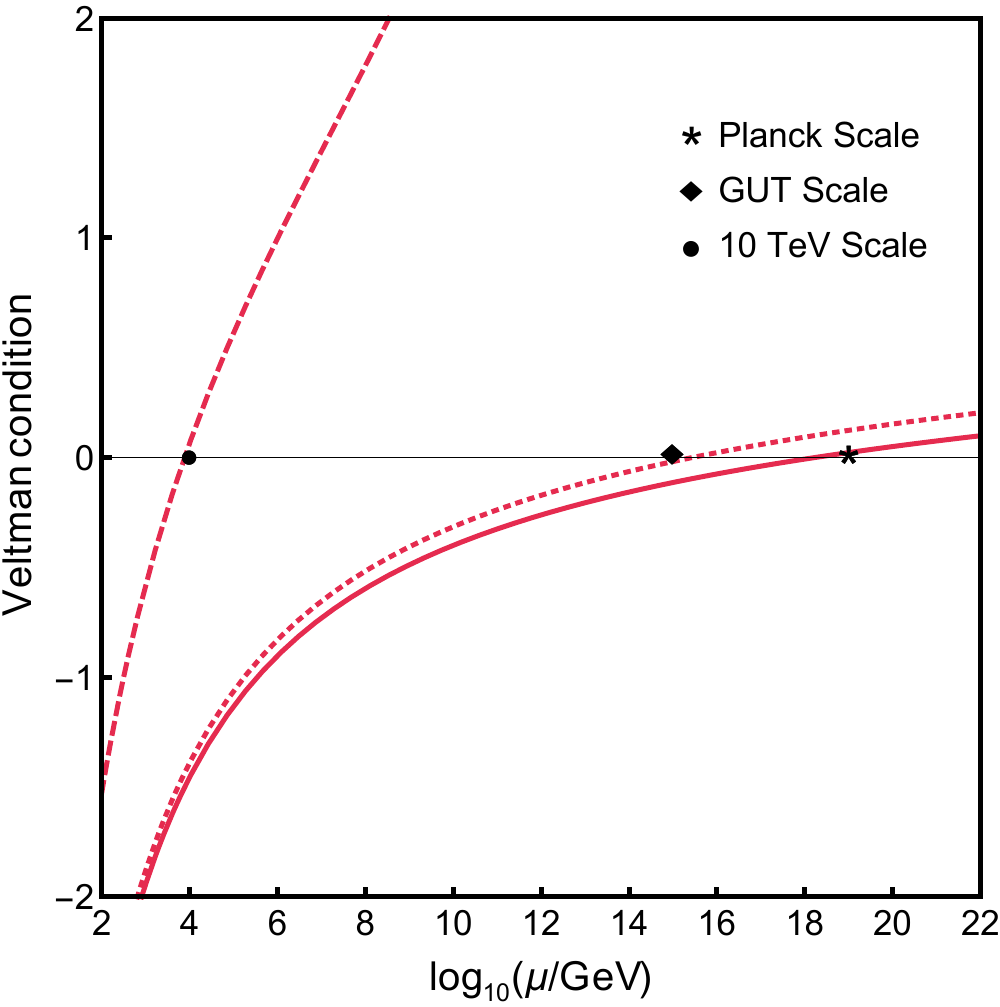}
		\caption{The running of the VC with three different choices of parameter spaces. The solid, dotted, and dashed curves represent the VC satisfied at the Planck scale, GUT scale and $10$\,TeV scale, respectively. The leptoquark-Higgs coupling $\gph=0.39$, $0.41$, and $0.8$ for the solid, dotted and dashed curves, respectively. All three curves are obtained for Higgs mass $M_h=125\,\gev$, top quark mass $M_t=173.3\,\gev$, strong coupling constant $\alpha_s=0.1185$, and NP `Yukawa type' couplings $\ylt=\yrt=0.5$ at the EW scale.} \label{fig:velt2} 
	\end{center}
\end{figure}

It should be noted that in exact supersymmetric theory the VC holds true for all orders and ensures the complete absence of quadratic divergences in the theory at any energy scale, whereas in a generic theory, the VC can be satisfied for a particular energy scale.
In the model we are considering in this paper, the SM Higgs mass gets an additional contribution from the extra charged scalar {\it i.e.,} the leptoquark at one-loop,
and the Higgs mass correction modifies as
\be\label{VClq}
\delta M^2_h= \frac{\Lambda^2}{16 \pi^2} \left( 3\lambda  + \frac{9}{4} g^2 + \frac{3}{4} g'^2 - 6 y^2_t+ 3\gph \right),
\ee
where $\gph$ is the leptoquark-Higgs coupling constant. Thus, the absence of quadratic divergence at a scale, say, $\mu_v$, at one-loop order can be found by
\be\label{VClq2}
 3\lambda (\mu_v) + \frac{9}{4} g^2(\mu_v) + \frac{3}{4} g'^2(\mu_v) - 6 y^2_t(\mu_v) +3\gph(\mu_v) =0.
\ee 
It should be noted that higher loop corrections to Eq.~\eqref{VClq2} will be further suppressed by $\mathcal{O}(1/(4\pi)^2)$ and hence will give a tiny modification to the energy scale $\mu_v$ at which the VC is satisfied.

In Fig.~\ref{fig:velt}, we have demonstrated the running of the couplings in Eq.~\eqref{VClq2}. We use two-loop running of the beta functions for all couplings. The plot is obtained by choosing: $M_h=125\,\gev$ for the Higgs mass and $\gph=0.25$ for the leptoquark-Higgs coupling constant and all other NP couplings are vanishing at the EW scale. The red band corresponds to the variation of the VC due to the uncertainty in top quark mass $M_t=173.3 \pm 0.87\, \gev$ \cite{mtop}.
It can be seen that Eq.~\eqref{VClq2} is only satisfied at $\sim 10^{22}~\gev$ for this particular choice of parameter space.

The VC demands that the quadratically divergent correction to the Higgs mass should be vanishing or at least small. In Fig.~\ref{fig:velt2}, we have shown three different scales at which the VC is exactly satisfied for particular choice of benchmark points of our model. The solid, dotted and dashed curves represent the VC satisfied at Planck scale, GUT scale and $10$\,TeV scale, respectively. The leptoquark-Higgs coupling 
$\gph=0.39$, $0.41$, and $0.8$ at the EW scale for the solid, dotted, and dashed curves, respectively. All three curves are obtained for Higgs mass $M_h=125\,\gev$, top quark mass $M_t=173.3\,\gev$, strong coupling constant $\alpha_s=0.1185$ and NP Yukawa couplings $\ylt=\yrt=0.5$ at the EW scale.

\begin{center}
	\begin{figure*}[!h]
		\begin{center}
			\includegraphics*[width=0.5\linewidth]{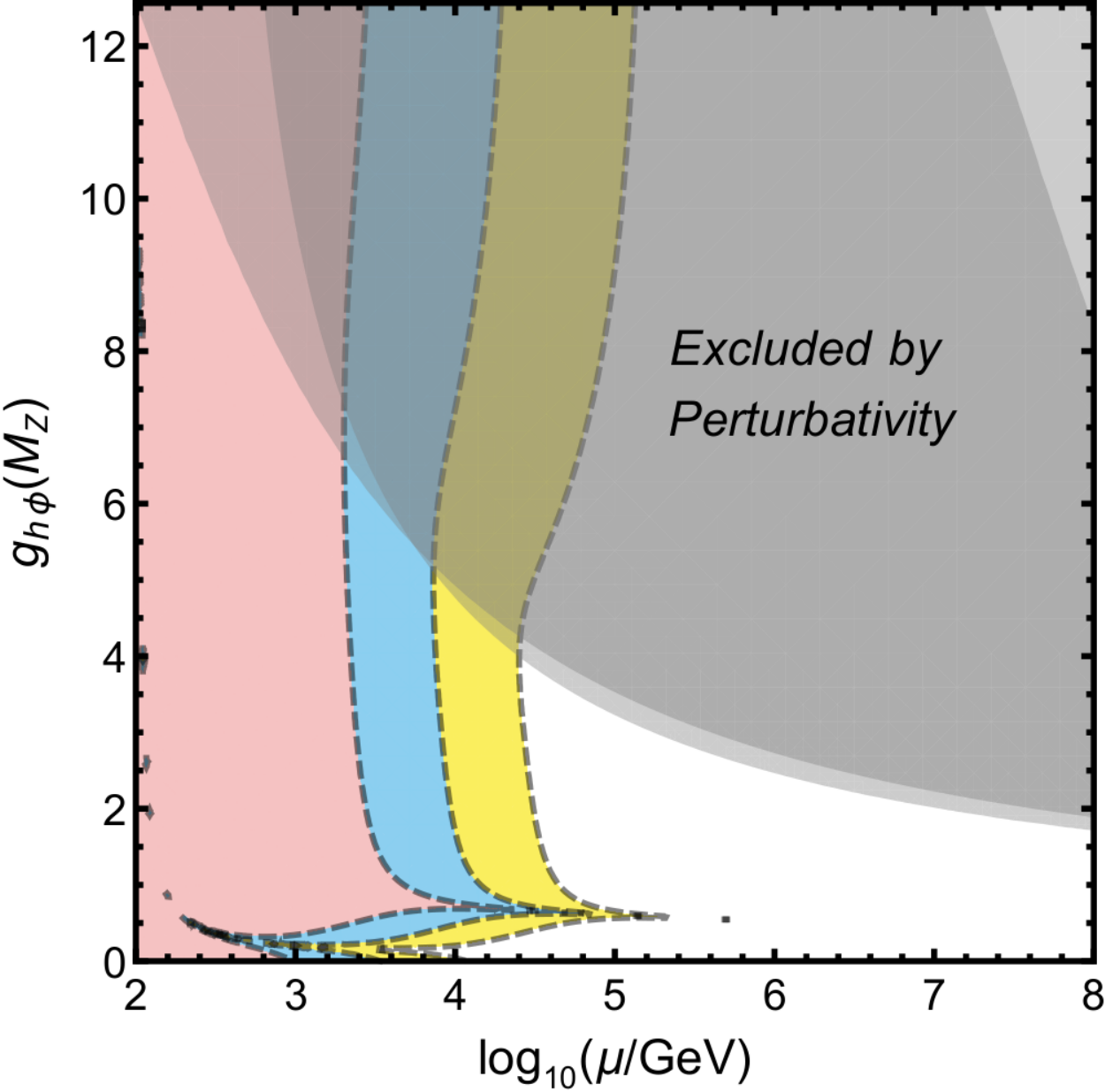}%
			\caption{The allowed region in the $\gph-\text{log}_{10}(\mu/\gev)$ plane for different values of the fine-tuning parameter $1/f$. The pink, blue and yellow regions denote no fine-tuning, up to 10\% fine-tuning and up to 100\% fine-tuning required for the model to remain valid up to the corresponding cutoff scale, respectively. The gray shaded region is excluded for perturbativity of the leptoquark-Higgs coupling $\gph$ and Higgs self-coupling $\lambda$. The plot is obtained for $\ylt=0.5$ and $\yrt=0.5$ values at the EW scale.} 
			\label{fig:fine}
		\end{center}
	\end{figure*}
\end{center}

Defining a fine-tune parameter $\displaystyle\frac{1}{f}\equiv \displaystyle\frac{|\delta M^2_h|}{M^2_h} $, in Fig.~\ref{fig:fine}, we highlight various different regions in the $\gph-\text{log}_{10}(\mu/\gev)$ plane for different values of the $f$ parameter. The pink, blue and yellow regions correspond to $f\ge1$, $1> f\ge0.1$ and $0.1> f\ge0.01$, respectively. These regions denote $\frac{1}{f} \leq 1\%$, $1 >\, \frac{1}{f} \leq10\%$ and $10 >\, \frac{1}{f} \leq 100\%$ fine-tuning,  respectively, required for the model to remain valid up to the corresponding cutoff scale. Even with $\sim 100\%$ fine-tuning, the validity scale of the theory remains within $\sim 10^6$ GeV.
Clearly, we see that higher fine-tuning allows a higher validity scale for the model.
On top of that, we also show the gray shaded region, which designates the disallowed region of parameter space by the demand for perturbativity of all couplings, especially, the Higgs self-coupling $\lambda$ and leptoquark-Higgs coupling $\gph$.

\section{Effect of leptoquark self-coupling}\label{phi4}
In the previous sections we discussed the simplified extension of SM with leptoquark $\phi$ where it interacts with the SM Higgs only via the coupling $\gph$ (see Eq.~\eqref{lag}). The introduction of an additional coupling, viz. the self-coupling for $\phi$ can change some of the conclusions which we intend to discuss here. The presence of leptoquark self-coupling term  $\lambda_{\phi} \phi^4$ in the Lagrangian in Eq.~\eqref{lag} modifies the running of leptoquark-Higgs coupling $\gph$ at one-loop order as follows:
\begin{align}
\label{eq:betagph2}
16\pi^2 \beta^{(1)}_{\gph}&= \frac{{g^\prime}^4}{3} +\gph \big(6 \lambda- \frac{9}{2} g^2- \frac{13}{6} {g^\prime}^2 -8 g_3^2 \big) +4 \gph^2 +4 \gph \Big({\yle}^2 +{\ylm}^2+{\ylt}^2 \Big)  \nn \\[2ex]  
	&+2 \gph \Big( {\yre}^2 +{\yrm}^2+{\yrt}^2  \Big) + y_t^2  \Big(6 \gph - 4 {\ylt}^2 - 4 {\yrt}^2 \Big ) + 16\gph \lambda_{\phi} .
\end{align}
\begin{center}
\begin{figure*}[h]
 \begin{center}
	\includegraphics*[width=0.5\linewidth]{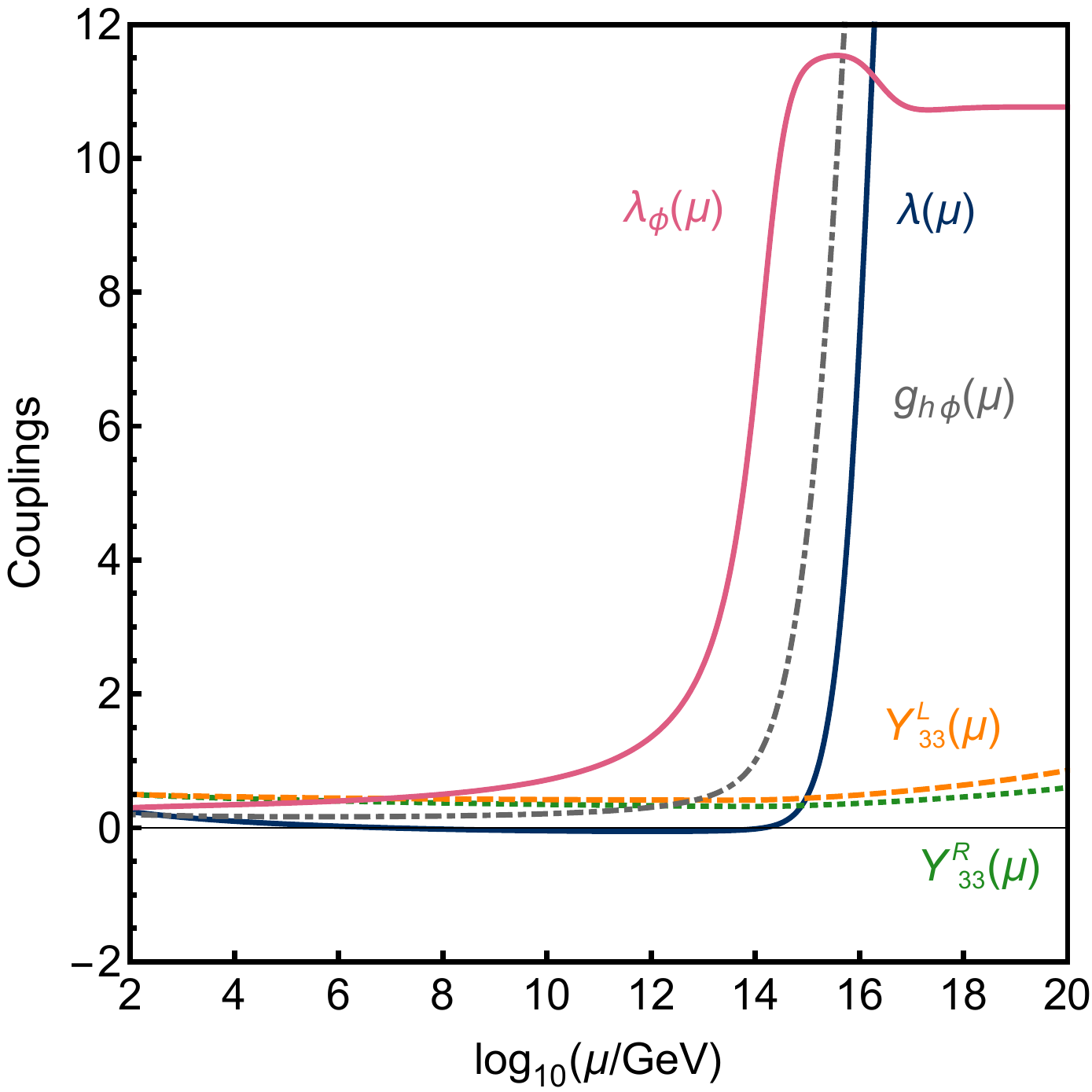}%
	 \caption{ The two-loop running of dimensionless coupling constants $\lambda_\phi$, $\lambda$, $\gph$, $\ylt$ and $\yrt$ is shown with the energy scale $\mu$ in pink (solid), blue (solid), gray (dashed-dotted), orange (dashed), and green (dotted) curves, respectively. The NP couplings are chosen as $\gph=0.2$, $Y_{22}^{\L,\R}=Y_{33}^{\L,\R}=0.5$ and $\lambda_\phi=0.3$ at the EW scale. The leptoquark-Higgs coupling $\gph$ hits the Landau pole around a scale $\sim 10^{15}\, \gev$. The leptoquark quartic coupling also takes a very large value around a scale $\sim 10^{13}\, \gev$.} 
	 \label{fig:quartic}
\end{center}
\end{figure*}
\end{center}

The one-loop beta function of leptoquark quartic coupling $\lambda_\phi$ is given by
\begin{align}
\label{eq:betalamphi}
16\pi^2 \beta^{(1)}_{\lambda_{\phi}}&= \frac{2}{27} {g^\prime}^4+\frac{4}{9}{g^\prime}^2 g_3^2+\frac{13}{6} g_3^4 -\lambda_{\phi}\left(\frac{4}{3}{g^\prime}^2+16 g_3^2\right)+2\gph^2 +28 {\lambda_{\phi}}^2 \nn \\
	&+8\lambda_{\phi} \Big( {\yle}^2 +{\ylm}^2+{\ylt}^2  \Big) +4\lambda_{\phi} \Big( {\yre}^2 +{\yrm}^2+{\yrt}^2  \Big) \nn \\
	&-4 \Big( {\yle}^4 +{\ylm}^4+{\ylt}^4  \Big)-2 \Big( {\yre}^4 +{\yrm}^4+{\yrt}^4  \Big) . 
\end{align}

Figure~\ref{fig:quartic} describes the running of NP couplings where the EW scale initial values are chosen as  $\gph=0.2$, $Y_{22}^{\L,\R}=Y_{33}^{\L,\R}=0.5$, and $\lambda_\phi=0.3$. It can be seen that $\gph$ hits the Landau pole at the energy scale $\sim 10^{15}\, \gev$ for the chosen parameter space.

We notice that $\lambda_\phi$ runs to negative values at the energy scale $\sim 10^{10}\,\gev$ for $\lambda_\phi(M_Z)$ $< 0.2$ and $Y_{22,\,33}^{\L,\R}(M_Z) > 0.75$ due to the dominance of terms in the last line of Eq.~\eqref{eq:betalamphi}. In such a situation the theory is unstable as the potential becomes unbounded from below in the $\phi$ direction. We also note that since the leptoquark quartic coupling $\lambda_\phi$ contributes negligibly to the running of Higgs self-coupling $\lambda$ only at two-loop level, the conclusions obtained in Sec.~\ref{stab} remain unchanged for the choice of parameter spaces where all the couplings in theory remain perturbative up to the Planck scale.

\section{Results}\label{res}

In this section, we summarize different bounds on the model arising from the perturbativity and stability analysis discussed in this paper: \\
$\bullet$ Perturbativity up to the Planck scale puts bound on the NP coupling when all the NP couplings are simultaneously turned on as,
$$\gph(M_Z)\le 0.55 \text{~~and~~} Y_{ii}^{\L,\R}(M_Z) \le 0.55;~~~~ i\in \{1,2,3\}.$$
However, as discussed before, due to the strong bound on $Y_{11}^{\L,\R}$ coupling \cite{Davidson:1993qk}, if a vanishing $Y_{11}^{\L,\R}$ value is assumed at the EW scale, the perturbativity bound is relaxed, which is given by
$$\gph(M_Z)\le 0.65 \text{~~and~~} Y_{ii}^{\L,\R}(M_Z) \le 0.65;~~~~ i\in \{2,3\}.$$
$\bullet$ The stability of the EW vacuum imposes a bound on the parameter space even within the perturbative region. Figure~\ref{fig:stability} shows that for $\gph(M_Z) \le 0.3$ there exists a cutoff scale (much before Planck scale) for the leptoquark model after which the Higgs potential becomes unstable. This bound on $\gph$ is insensitive to $Y_{ii}^{\L,\R}(M_Z)$ values (within the perturbative regime), however, it has a significant impact from top quark mass $M_t$. Hence for the current measurement of $M_t$, the combined bound on $\gph$ from the perturbativity and vacuum stability is $$0.3< \gph(M_Z)\le 0.65\, .$$ Thus, the constraints from both the vacuum stability and perturbativity give a very predictive value for the NP couplings, which can be tested in collider experiments.

\section{Phenomenology}\label{pheno}

The introduction of charged scalar field $\phi$ affects the stability of the Higgs potential via its coupling to the Higgs boson, {\it i.e.,} $g_{h\phi}$, and also the perturbativity of the Higgs self-coupling $\lambda$ via beta functions. To prove the existence of such an extra scalar, we have to look for it at colliders.
It would be interesting if the LHC or a future collier can test the signatures of such a charged scalar. The fractional electromagnetic charge of the particle, not participating in the EWSB (not getting a vev) and its tree level decays to lepton-quark flavor violating modes make such a field much more interesting.  In this section we qualitatively discuss the various possible decays and phenomenologically viable signatures that we are going to explore in our next study \cite{pbrm}.

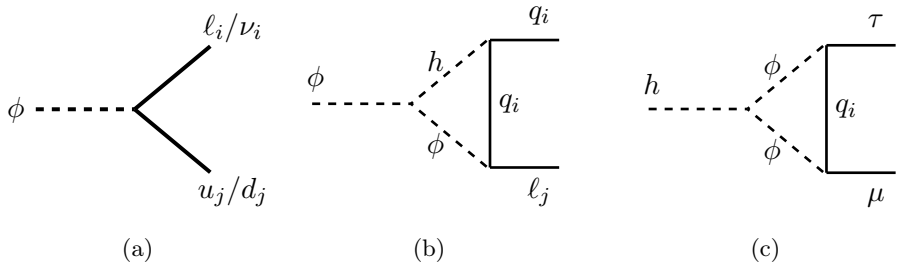
\begin{figure}[!h]
\begin{center}
\mbox{\subfigure[]{\begin{tikzpicture}[line width=1.5 pt, scale=1.3]
	\draw[fermion] (-40:1)--(0,0);
	\draw[fermion] (40:1)--(0,0);
	\draw[scalar] (180:1)--(0,0);
	\node at (-40:1.3) {$u_j/d_j$};
	\node at (40:1.3) {$\ell_i/\nu_i$};
	\node at (180:1.2) {$\phi$};
\end{tikzpicture}}
\subfigure[]{\begin{tikzpicture}[line width=1. pt, scale=1.3]
	\draw[scalar] (-40:1)--(0,0);
	\draw[scalar] (40:1)--(0,0);
	\draw[scalar] (180:1)--(0,0);
	\draw[fermion] (0.8,0.65)--(0.8,-0.65) node at (1,0.) {$q_i$} ;
	\node at (-60:0.5) {$\phi$};
	\node at (60:0.5) {$h$};
	\node at (-15:-1) {$\phi$};
	\draw[fermion] (0.8,0.65)--(1.5,0.65) node at (1.3,0.9) {$q_i$};
	\draw[fermion] (0.8,-0.65)--(1.5,-0.65) node at (1.3,-0.9) {$\ell_j$} ;
 \end{tikzpicture}}
\hskip 20pt \subfigure[]{\begin{tikzpicture}[line width=1. pt, scale=1.3]
	\draw[scalar] (-40:1)--(0,0);
	\draw[scalar] (40:1)--(0,0);
	\draw[scalar] (180:1)--(0,0);
	\draw[fermion] (0.8,0.65)--(0.8,-0.65) node at (1,0.) {$q_i$} ;
	\node at (-60:0.5) {$\phi$};
	\node at (60:0.5) {$\phi$};
	\node at (-15:-1) {$h$};
	\draw[fermion] (0.8,0.65)--(1.5,0.65) node at (1.3,0.9) {$\tau$};
	\draw[fermion] (0.8,-0.65)--(1.5,-0.65) node at (1.3,-0.9) {$\mu$} ;
 \end{tikzpicture}}}
\caption{Some examples of NP diagrams due to the presence of leptoquark $\phi$ where (a), (b) show the lepton-quark flavor violating
decay of $\phi$ and (c) shows lepton-quark flavor violating decay of Higgs boson. These decays may have important collider signatures (see the text for details).}\label{dia:pheno}
\end{center}
\end{figure}

The vacuum stability and perturbativity bounds suggest that a moderate value of $g_{h\phi} \lesssim 0.5$ is phenomenologically viable. Earlier study on this model and the bounds from different flavor constraints suggest that $m_{\phi}\geq 1$ TeV \cite{Bauer:2015knc}, and the recent experimental bounds on the search of scalar leptoquarks in collider suggest that $m_{\phi}\geq 1.2$ TeV \cite{Aaboud:2016qeg}, but such bounds are model dependent. These automatically make $\phi$ rather heavy, which can decay to a quark and lepton via $Y_{ii}^{\L, \R}$ couplings at the tree level. Loop decays with nonzero off-diagonal terms of $Y_{ii}^{\L, \R}$ can induce not only quark and lepton flavor but also generation violating decays; viz., we can observe the decay of $\phi$ into $u, \mu$, etc., as can be seen from Fig.~\ref{dia:pheno}(a). A similar graph can also promote the Higgs boson decaying to $\mu\tau$ as we can see from Fig.~\ref{dia:pheno}(c). Some excess in $h \to \mu \tau $ decay has already been observed by CMS \cite{Khachatryan:2015kon}, and the presence of this kind of coupling \cite{Dorsner:2015mja} may explain such observed excess.


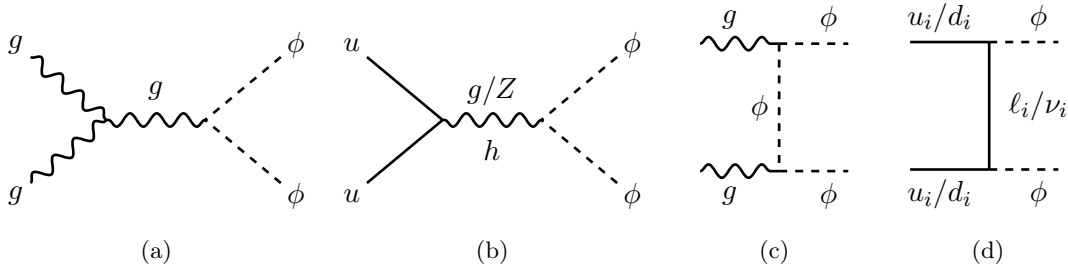
\begin{figure}[!h]
\begin{center}
\mbox{ \hspace{-0.5 cm}\subfigure[]{\begin{tikzpicture}[line width=1. pt, scale=1.3]
	\draw[vector] (-140:1)--(0,0);
	\draw[vector] (140:1)--(0,0);
	\draw[vector] (0:1)--(0,0);
	\node at (-140:1.2) {$g$};
	\node at (140:1.2) {$g$};
	\node at (.5,.3) {$g$};	
\begin{scope}[shift={(1,0)}]
	\draw[scalar] (-40:1)--(0,0);
	\draw[scalar] (40:1)--(0,0);
	\node at (-40:1.2) {$\phi$};
	\node at (40:1.2) {$\phi$};	
\end{scope}
\end{tikzpicture}}
 \subfigure[]{\begin{tikzpicture}[line width=1. pt, scale=1.3]
	\draw[fermion] (-140:1)--(0,0);
	\draw[fermion] (140:1)--(0,0);
	\draw[vector] (0:1)--(0,0);
	\node at (-140:1.2) {$u$};
	\node at (140:1.2) {$u$};
	\node at (.5,.3) {$g/Z$};
	\node at (.5,-.3) {$h$};
\begin{scope}[shift={(1,0)}]
	\draw[scalar] (-40:1)--(0,0);
	\draw[scalar] (40:1)--(0,0);
	\node at (-40:1.2) {$\phi$};
	\node at (40:1.2) {$\phi$};	
\end{scope}
\end{tikzpicture}}
  \hspace{0.25 cm}
 \subfigure[]{\begin{tikzpicture}[line width=1 pt, scale=1.3]
	\draw[vector] (0.,0.65) node at (0.3,0.9) {$g$}--(0.8,0.65);
	\draw[scalar] (0.8,0.65)--(1.5,0.65) node at (1.3,0.9) {$\phi$};
	\draw[scalar] (0.8,0.65)node at (0.6,0) {$\phi$}--(0.8,-0.65);
	\draw[vector] (0.,-0.65) node at (0.3,-0.9) {$g$}--(0.8,-0.65);
	\draw[scalar] (0.8,-0.65)--(1.5,-0.65) node at (1.3,-0.9) {$\phi$};
 \end{tikzpicture}}
  \hspace{0.25 cm}
\subfigure[]{\begin{tikzpicture}[line width=1 pt, scale=1.3]
	\draw[fermion] (0.,0.65) node at (0.3,0.9) {$u_i/d_i$}--(0.8,0.65);
	\draw[scalar] (0.8,0.65)--(1.5,0.65) node at (1.3,0.9) {$\phi$};
	\draw[fermion] (0.8,0.65)node at (1.3,0) {$\ell_i/\nu_i$}--(0.8,-0.65);
	\draw[fermion] (0.,-0.65) node at (0.3,-0.9) {$u_i/d_i$}--(0.8,-0.65);
	\draw[scalar] (0.8,-0.65)--(1.5,-0.65) node at (1.3,-0.9) {$\phi$};
 \end{tikzpicture}}}
\caption{Feynman diagrams showing different production modes of $\phi\phi$; (a), (c) from gluon fusion and (b), (d) from quark annihilation
process, at the LHC}\label{dia:pheno2}
\end{center}
\end{figure}

\begin{figure}[!h]
\begin{center}
\mbox{\subfigure[]{\includegraphics[width=0.37\linewidth, angle=-90]{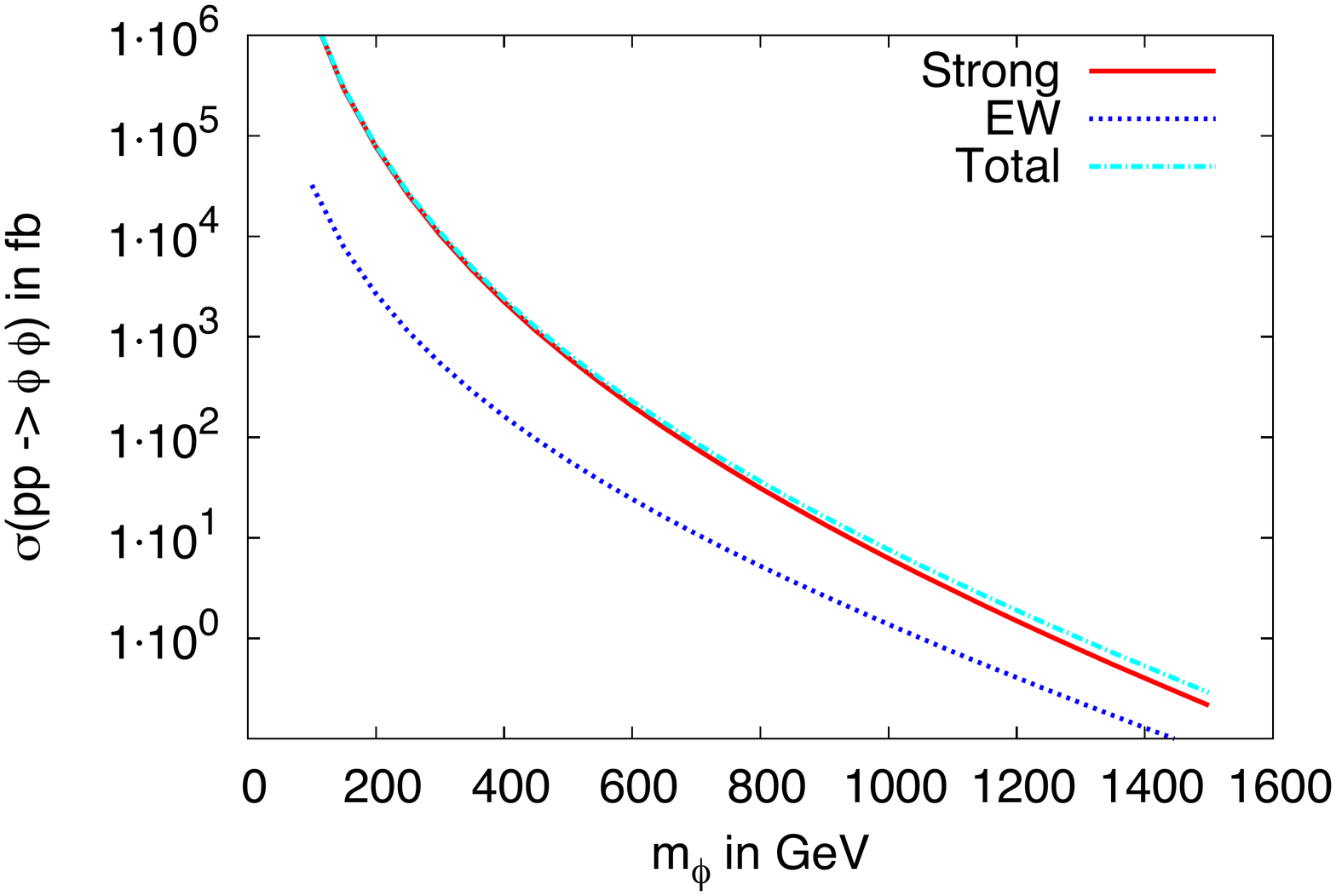}}
\hspace*{-5mm}\subfigure[]{\includegraphics[width=0.37\linewidth, angle=-90]{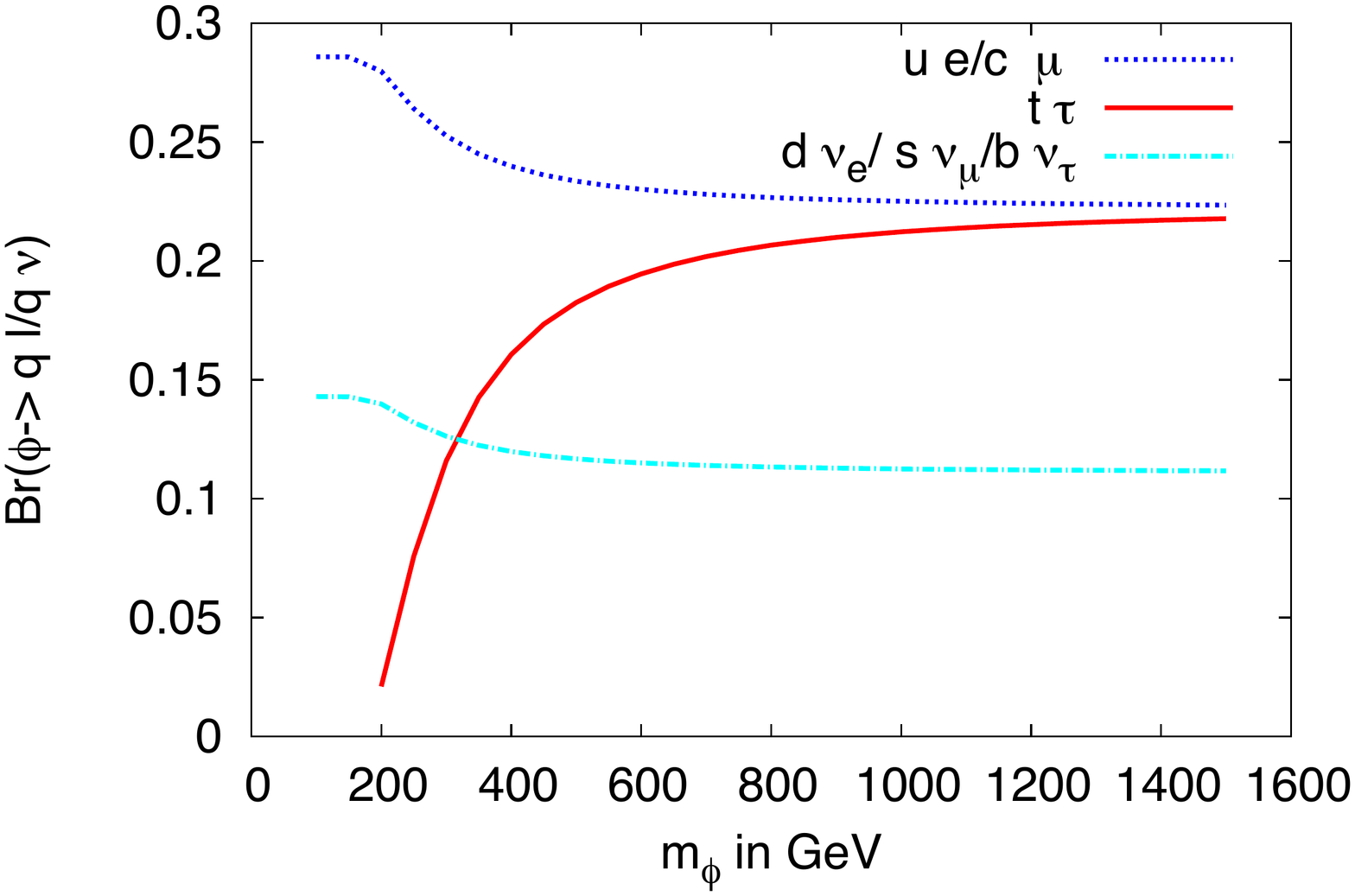}}}
\caption{The production cross-section of $\phi \phi$ at the LHC at 14 TeV $E_\text{CM}$ vs the leptoquark mass is shown in panel (a) and the lepton-quark flavor violating decay branching fraction of $\phi$ vs leptoquark mass is depicted in panel (b). The plots are obtained for $\gph=0.5$ and $Y_{ii}^{\L,\R}=0.5$ at the EW scale.}\label{Proddecay}
\end{center}
\end{figure}

\begin{figure}[!h]
\begin{center}
\begin{tikzpicture}[line width=1 pt, scale=1.3]
	\draw[fermion] (0.,0.65) node at (0.3,0.9) {$u$}--(1.5,0.65) node at (1.3,0.9) {$\mu$};
	\draw[scalar] (0.8,0.65)node at (0.6,0) {$\phi$}--(0.8,-0.65);
	\draw[fermion] (0.,-0.65) node at (0.3,-0.9) {$d$}--(1.5,-0.65) node at (1.3,-0.9) {$\mu$};
 \end{tikzpicture}
\caption{Production of $\mu^+\mu^-$ via the $t$-channel exchange of $\phi$.}\label{dia:pheno3}
\end{center}
\end{figure}
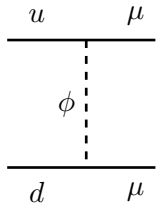

Production of $\phi$ at the LHC can be quite easier because of its color charge. Figure~\ref{dia:pheno2}
shows different production mechanisms for a $\phi$ pair final state. In the $s$ channel, photon and $Z$ boson contribute due to a nonzero hypercharge of $\phi$, but the gluon contributes heavily due to the color charge of the leptoquark and also because of larger gluon flux at the LHC. There are also $t$-channel contributions coming from $\phi$ and leptons, as can be seen from Figs.~\ref{dia:pheno2}(c) and (d). 

Figure~\ref{Proddecay}(a) shows the variation of the $\phi$ pair production cross-section with the leptoquark mass at the LHC with 14 TeV center-of-mass energy ($E_\text{CM}$) for $\gph=0.5$ and $Y_{ii}^{\L,\R}=0.5$ at the EW scale. The tree-level cross-sections have been calculated by implementing the model in CalcHEP$\_$3.6.23  \cite{calchep}.  We can see that EW cross-section, though subdominant compared to the strong production, can be very useful for a lower mass of $\phi$. For 1 TeV $\phi$, the cross-section reaches to around $10$ fb at the LHC with 14 TeV $E_\text{CM}$, for a Parton Distribution Function (PDF) choice of CTEQ6L \cite{6teq6l} and scale of $\sqrt{\hat{S}}$. Because to the larger cross-section, both the quark and lepton flavor violating signatures can be probed at the LHC with 14 TeV $E_\text{CM}$. 

Figure~\ref{Proddecay}(b) shows the variation of the decay branching fraction with the leptoquark mass for lepton-quark flavor- as well as number violating decays for $\gph=0.5$ and $Y_{ii}^{\L,\R}=0.5$ at the EW scale. For a lower mass range, the decay branching fractions to $u\, e$ and $c\, \mu$ are rather large, $25\%-30\%$, compared to the decays into the third generations, {\it i.e.,} $t \, \tau$, or into the down type quarks and neutrinos which stay around $\lesssim 15\%$. For a TeV mass leptoquark, its decay branching fraction in $t \, \tau$ increases to $\sim 23\%$.  However, assuming $Y_{11}^{\L,\R}=0$, which satisfies the flavor data \cite{Davidson:1993qk}, the decays of the leptoquark to channels $u e$ and $d \nu_e$ are forbidden. Hence, in this scenario, the branching fractions for second and third generation quarks, leptons i.e., ($c \,\mu, t \,\tau$) and ($s \,\nu_{\mu}, b \,\nu_{\tau}$), are increased and are given by $\sim 33\% \, \rm{and} \,17\%$, respectively. A final state consisting of $2t +2\tau$ can lead to $2b+4j+2\tau $ and  $2b +2\ell + 2\tau +\ptmiss$ final states, which can probe the lepton-quark flavor violating decays. On the other hand decays to the light quark can be searched for where a pair-  produced $\phi$ leads to $\rm{dijet}+ 1\ell + \ptmiss$ via two different types of decays.  For a TeV mass $\phi$, its decay to quarks plus charged leptons is $66.1\%$ and to quarks plus neutrinos is $33.9\%$. The decays of $\phi$ into both the first two generations and the third generation will be useful at the LHC to search for  lepton-quark flavor violating decays of the leptoquark  predicted by this model. 

Apart from the above-mentioned channels, up and down quark fusion via $\phi$ can give rise to a $\mu^+\mu^-$ final state. It can be see from Fig.~\ref{dia:pheno3} that this is a $t$-channel process and thus such an observation will not produce a $Z$ mass peak in the $\mu^+\mu^-$ invariant mass distribution. The interference of such a diagram with normal SM $s$-channel modes can be tested at the LHC while looking for different differential distributions of the $\mu^+\mu^-$ final state.

\section{Conclusions} \label{concl}
In this article, we explore the vacuum stability constraints for the SM extension with a colored electromagnetically charged scalar: the leptoquark. The introduction of this scalar is motivated in order to explain some anomalies in $B$ decays and the muon $g-2$ discrepancy. We see that the requirement of all the couplings to be perturbative until the Planck scale imposes bounds on the NP couplings at the EW scale, viz. $\gph(M_Z)\le 0.65 \text{~and~} Y_{ii}^{\L,\R}(M_Z) \le 0.65;~i\in \{2,3\}$, when we assume vanishing $Y_{11}^{\L,\R}$ couplings at the EW scale due to the constraints arising from flavor data \cite{Davidson:1993qk}, while the bounds from vacuum stability demands that $\gph(M_Z)\ge 0.3$ in order to have the Higgs potential (meta)stable till Planck scale. These bounds are very sensitive to the values of top quark mass. 
The perturbativity of all dimensionless couplings is also studied for various choices of the parameter space, and we find that high values of NP couplings at the EW scale often lead to a situation where the Higgs self-coupling $\lambda$ and/or leptoquark-Higgs coupling $\gph$ hit(s) the Landau pole.

We also address the issue of fine-tuning, and the presence of the leptoquark $\phi$ certainly reduces the SM Higgs mass fine-tuning by contributing at one-loop Higgs mass. We define the measure of fine-tuning by the VC \cite{vltmn}. Allowing $100\%$ fine-tuning, the scale until which the theory remains valid is $\mu \lesssim 10^{6}$ GeV.

The allowed moderate values of the NP couplings $\gph$ and $Y_{ii}^{\L,\R}$ at the EW scale provide us a scope to measure them at current and future colliders. The obvious way to measure them is via producing them and observing their decay products.  We also discuss a different production mechanism for the $\phi$ pair at the LHC. The hard scattering cross-section suggests that such a $\mathcal{O}\,$(TeV) $\phi$ can be easily probed at the LHC with 14 TeV $E_\text{CM}$. The distinguishing signature for such a scalar is that it violates both the lepton and baryon number  at the same time. The effect of $\phi$ in the normal Drell-Yan process is also worth observing. The effect of the leptoquark in lepton flavor violating loop decays of the SM Higgs boson is an important observable which can further constrain the model when tested against experimental data.

\section*{Acknowledgments }
R.M. thanks Rahul Sinha for useful discussions and encouragement. R.M. also thanks Romesh Kaul and Biswarup Mukhopadhyaya for discussing some critical issues regarding stability analysis. P.B. wants to acknowledge The Institute of Mathematical Sciences, Chennai for the visit during the initial part of the project.

\end{document}